\documentclass[12pt,letter,useAMS,natbib]{mn2e}
\usepackage{graphicx,amssymb,amsmath}
\usepackage{txfonts}
\usepackage{natbib_jrm}
\usepackage{tikz}
\usetikzlibrary{fit,shapes.misc}
\bibpunct[, ]{(}{)}{;}{a}{}{,}
\def \aj {AJ}
\def \mnras {MNRAS}
\def \pasp {PASP}
\def \apj {ApJ}
\def \apjs {ApJS}
\def \apjl {ApJL}
\def \aap {A\&A}
\def \aaps {A\&A Suppl.}
\def \nat {Nature}
\def \araa {ARAA}

\newcommand{\sloang} {$g^{\prime}\ $}
\newcommand{\sloanr} {$r^{\prime}\ $}
\newcommand{\sloani} {$i^{\prime}\ $}

\def\lesssim{\mathrel{\hbox{\rlap{\hbox{\lower4pt\hbox{$\sim$}}}\hbox{$<$}}}}
\def\gtrsim{\mathrel{\hbox{\rlap{\hbox{\lower4pt\hbox{$\sim$}}}\hbox{$>$}}}}
\newcommand{\halpha} {$\mathrm{H\alpha}\,$}

\long\def\symbolfootnote[#1]#2{\begingroup%
\def\thefootnote{\fnsymbol{footnote}}\footnote[#1]{#2}\endgroup} 
\newcommand\marktopleft[1]{%
    \tikz[overlay,remember picture] 
        \node (marker-#1-a) at (0,1.5ex) {};%
}
\newcommand\markbottomright[1]{%
    \tikz[overlay,remember picture] 
        \node (marker-#1-b) at (0,0) {};%
    \tikz[overlay,remember picture,thick,dashed,inner sep=3pt]
        \node[draw,rectangle,fit=(marker-#1-a.center) (marker-#1-b.center)] {};%
}

\begin{document}

%%%%%%%%%%%%%%%%%%%%%%%%%%%%%%%
%Title Page                                                                                        %
%%%%%%%%%%%%%%%%%%%%%%%%%%%%%%%
\title[The Progenitor of SN 2008bk Revisited]{A new precise mass for the progenitor of the Type IIP SN~2008bk\thanks{Based on observations collected for program 087.D-0587 at the European Organisation for Astronomical Research in the Southern Hemisphere, Chile}\thanks{Based on observations obtained for program GS-2011B-Q-21 at the Gemini Observatory, which is operated by the Association of Universities for Research in Astronomy, Inc., under a cooperative agreement with the NSF on behalf of the Gemini partnership: the National Science Foundation (United States), the National Research Council (Canada), CONICYT (Chile), the Australian Research Council (Australia), Minist\'{e}rio da Ci\^{e}ncia, Tecnologia e Inova\c{c}\~{a}o (Brazil) and Ministerio de Ciencia, Tecnolog\'{i}a e Innovaci\'{o}n Productiva (Argentina)}\thanks{Based on observations made with the NASA/ESA Hubble Space Telescope, which is operated by the Association of Universities for Research in Astronomy, Inc., under NASA contract NAS 5-26555. These observations are associated with program GO-12262.}}
\author[Maund et al.]{Justyn R. Maund$^{1,2}$\thanks{j.maund@qub.ac.uk}, Seppo Mattila$^{3}$, Enrico Ramirez-Ruiz$^{4}$ \& John J. Eldridge$^{5}$\\ 
$^{1}\,$ Astrophysics Research Centre, School of Mathematics and Physics, Queen's University Belfast, Belfast, BT7 1NN, Northern Ireland\\
$^{2}\,$ Royal Society Research Fellow\\
$^{3}\,$  Finnish Centre for Astronomy with ESO (FINCA), University of Turku, 
V\"{a}is\"{a}l\"{a}ntie 20, FI-21500 Piikki\"{o}, Finland\\
$^{4}\,$ Department of Astronomy \& Astrophysics, University of California, Santa Cruz, 95064, U.S.A.\\
$^{5}\,$ Department of Physics, University of Auckland, Private Bag 92019, Auckland, New Zealand}

\maketitle
\begin{abstract}
The progenitor of the Type IIP SN~2008bk was discovered in pre-explosion $g^{\prime}r^{\prime}i^{\prime}IYJHK_{s}$ images, acquired with European Southern Observatory Very Large Telescope FORS, HAWK-I and ISAAC instruments and the Gemini GMOS-S instrument.  The wealth of pre-explosion observations makes the progenitor of this SN one of the best studied, since the detection of the progenitor of SN~1987A. Previous analyses of the properties of the progenitor were hampered by the limited quality of the photometric calibration of the pre-explosion images and the crowded nature of the field containing the SN.  We present new late-time observations of the site of SN~2008bk acquired with identical instrument and filter configurations as the pre-explosion observations, and confirm that the previously identified red supergiant star was the progenitor of this SN and has now disappeared.  Image subtraction techniques were used to conduct precise photometry of the now missing progenitor, independently of blending from any nearby stars.  The nature of the surrounding stellar population and their contribution to the flux attributed to the progenitor in the pre-explosion images are probed using $HST$ $WFC3$ $UVIS/IR$ observations.  In comparison with MARCS synthetic spectra, we find the progenitor was a highly reddened RSG with luminosity $\log (L/L_{\odot})=4.84^{+0.10}_{-0.12}$, corresponding to an initial mass of $M_{init}=12.9^{+1.6}_{-1.8}M_{\odot}$.  The temperature of the progenitor was hotter than previously expected for RSGs ($T\sim 4330K$), but consistent with new temperatures derived for RSGs using SED fitting techniques.  We show that there is evidence for significant extinction of the progenitor, possibly arising in the CSM; but that this dust yields a similar reddening law to dust found in the ISM ($E(B-V)=0.77$ with $R_{V}=3.1$).  Our improved analysis, which carefully accounts for the systematics, results in a more precise and robust mass estimate, making the progenitor of SN 2008bk the  most well understood progenitor of a Type IIP SN from pre-explosion observations.
\end{abstract}

%%%%%%%%%%%%%%%%%%%%%%%%%%%%%%%
%Introduction   
%Introduction 
%Introduction 
%Introduction 
%Introduction 
%Introduction 
%%%%%%%%%%%%%%%%%%%%%%%%%%%%%%%
\section{Introduction}
\label{sec:intro}

All single stars with initial masses in the range $8 \lesssim M_{init}\lesssim
25-30M_{\odot}$ are predicted to end their lives as Red
Supergiants (RSGs).  These stars are expected to end their evolution
with a core-collapse induced supernova (SN) explosion; with the
massive H-rich envelope, retained by the RSG, giving rise to the
characteristic light curve plateau and H-dominated spectra of Type IIP
SNe.\\
Although there have been efforts to infer the properties of the
progenitor star through interpreting the observed characteristics of
the subsequent SN \citep[e.g.][]{2008A&A...491..507U,2013MNRAS.433.1745D}, the
direct observation of the progenitor in fortuitous pre-explosion
images has provided direct constraints on the properties of the
progenitors independently of models of how SNe explode
\citep[for a review see][]{2009ARA&A..47...63S}.  The greatest success in the direct
detection of the progenitors has been for the Type IIP SNe, originating
from RSGs \citep{2008arXiv0809.0403S}.  An
analysis of the population of observed and undetected
progenitors with mass constraints, showed that Type IIP SNe appear
to only arise from progenitor stars in the mass range $8 <
M_{init}\lesssim 16_{\odot}$, implying that RSGs evolving from more massive stars
were either exploding as a different type of SN or, perhaps, not
exploding as observable SNe.  \citet{2008arXiv0809.0403S} referred to
the observed deficit of massive RSG progenitors as the ``RSG
problem''.  

Recently, \citet{2012MNRAS.419.2054W} and
\citet{2012ApJ...759...20K} have explored the role of dust in the
circumstellar mediums (CSMs) of RSGs progenitors, in addition to
foreground Galactic and host reddening components, in dimming the
most massive and luminous progenitors ($\geq 15 M_{\odot}$), causing the masses determined
for these RSGs to be systematically under-estimated \citep{2012MNRAS.419.2054W}.  The determination
of the masses of the progenitors stars of Type IIP SNe have, however,
relied on observations of fortuitous nature.  These observations are often shallow and
have limited wavelength coverage.  The direct assessment of the
roles of multiple reddening components have, therefore, been
undermined by the degeneracies between the different parameters (in
particular reddening and temperature) which cannot generally be broken with the
available data and which, therefore, have serious consequences for the determination of
the luminosity and mass of the progenitors. 

The application of Spectral Energy Distribution (SED) techniques to progenitor stars detected in pre-explosion images provides a robust method to evaluate the degeneracies between parameters and quantitatively explore the consequences of additional reddening components.  
\citet{2013arXiv1302.7152M} showed, using MARCS synthetic spectra, that previous observations of the progenitors of Type IIP SNe were not consistent with large amounts of CSM dust with composition and reddening law significantly different to dust in the interstellar medium (ISM).  This could imply that dust may not be a solution to the RSG problem.

SN~2008bk was discovered by \citet{2008CBET.1315....1M} on 2008 Mar 25.14, which \citet{2008ATel.1448....1L} subsequently placed $9.2\arcsec$E and $126.4\arcsec$N of the centre of the host galaxy NGC 7793.  \citeauthor{2008ATel.1448....1L} also identified a possible red star close to the position of SN~2008bk in the pre-explosion images, for which \citet{2008ATel.1464....1M} reported European Southern Observatory (ESO) Very Large Telescope (VLT) ISAAC $J$ and $K_{S}$ photometry.  Previously \citet{2008ApJ...688L..91M} and \citet{2012AJ....143...19V} have presented the full identification and parameterisation of the progenitor, independently determining the initial mass of the progenitor to be $8.5\pm1$ and $8-8.5M_{\odot}$, respectively.  \citet{2010arXiv1011.5494M} later reported the subsequent disappearance of the progenitor, confirming the identity of the red supergiant observed in pre-explosion images.

Due to the wealth of multi-wavelength pre-explosion observations, the progenitor of SN~2008bk presents the opportunity to provide the best constraints for the progenitors of any SN since SN~1987A \citep{gil87a}.  Here we present new late-time observations of the site of SN~2008bk, acquired with the Gemini South Telescope, the European Southern Observatory Very Large Telescope (VLT) and the Hubble Space Telescope to open a new window onto the nature of the progenitor object.  

In Section \ref{sec:obs} we present the new late-time observations of the site of SN~2008bk and their reduction.  The results of these new observations are presented in Section \ref{sec:res}, and the outcome of our extensive analysis is represented in Section \ref{sec:ana}.  In Section \ref{sec:disc} we present a discussion of our results and our conclusions.
%%%%%%%%%%%%%%%%%%%%%%%%%%%%%%%
%Observations                                                                  
%Observations
%Observations
%Observations sec:obs
%Observations
%Observations
%%%%%%%%%%%%%%%%%%%%%%%%%%%%%%%
\section{Observations \& Data Reduction}
\label{sec:obs}
The details of the pre-explosion, post-explosion and late-time images used here are presented in Table \ref{tab:obs}.

\subsection{VLT FORS, HAWK-I and ISAAC Observations}
The pre-explosion ESO VLT FORS, HAWK-I and ISAAC observations were
previously presented by \citet{2008ApJ...688L..91M} (the near-infrared
images were also used by \citealt{2012AJ....143...19V}).  Late-time VLT observations of the site of SN~2008bk were acquired with almost identical instrument configurations as the pre-explosion observations, as part of program 087.D-0587 (PI Maund); these late-time VLT observations were acquired under photometric conditions.  The
near-infrared HAWK-I and ISAAC late-time observations were acquired
using identical instrument configurations as the pre-explosion
observations $\sim \mathrm{3.4 yrs}\,$ after explosion.  The
late-time FORS observations were acquired with a close approximation
of the instrument configuration used for the pre-explosion
observations: the pre-explosion observations utilised the FORS1
instrument with the Tektronix detector providing a pixel scale of
$0.2\arcsec$, whereas the late-time observations used the newly
designated FORS instrument (originally FORS2) with the red optimised
MIT detectors providing a pixel scale $0.252\arcsec$
\citep{1998Msngr..94....1A}.  The data were reduced using the
corresponding ESO
pipelines\footnote{http://www.eso.org/sci/software/pipelines/}.  

The
FORS data were bias and flatfield corrected in the standard fashion.
The infra-red HAWK-I and ISAAC data were bias, flat corrected and sky subtracted (using interleaved sky offset frames) before combination.  Photometry of the late-time frames was conducted
using {\sc iraf} and the Point Spread Function (PSF) fitting routines in the
{\tt daophot} package.  The data were photometrically calibrated with
zeropoints, colour terms and extinction coefficients derived from
observations of standard fields, acquired on the same nights.  

The pre-explosion and late-time FORS
observations were conducted with the Bessell $BVI$ filters, but the
corresponding zeropoints and colour terms were used to place the
photometry in the Johnson-Cousins $BVI$ photometric system.  The
HAWK-I and ISAAC data were similarly calibrated using observations of
infrared standard stars in the $YJHK_{S}$ filters.  The $JHK_{S}$
zeropoints were checked against photometry of bright stars in the
late-time images that were also in the 2MASS point source catalogue,
with a corresponding dispersion of $0.03$, $0.08$ and $0.1$
magnitudes, respectively.  The pre-explosion photometry was
bootstrapped to the photometric calibration of the late-time images.

\subsection{Gemini GMOS-S Observations}
The pre-explosion Gemini GMOS-S observations were previously presented
by \citet{2012AJ....143...19V}.  We attempted to re-identify the data
used by \citeauthor{2012AJ....143...19V} in the Gemini
archive\footnote{http://www3.cadc-ccda.hia-iha.nrc-cnrc.gc.ca/gsa/},
according to the description of observations provided by
\citeauthor{2012AJ....143...19V}.  The data were reduced in the
standard fashion using {\sc iraf} and the specific {\tt gemini} data
reduction package\footnote{http://www.gemini.edu/node/10795}.  The
data were bias and flatfield corrected, and the individual chips for
each exposure were combined to produce mosaiced images.  The mosaic
images were coadded to produced deep images for each filter.  The
pre-explosion observations were conducted in filters close to the Sloan Digital Sky
Server (SDSS) {\it g'r'i'} filters, but were conducted under
non-photometric conditions.
 
Late-time observations Gemini GMOS-S were conducted with an identical
instrument setup, under improved seeing and photometric
conditions compared to the pre-explosion observations, $\sim3.5\,\mathrm{ yrs}\,$ post-explosion as part of program GS-2011B-Q-21 (PI Maund).  The data were
reduced in the same fashion as the pre-explosion Gemini observations.
Photometry of the images was conducted using IRAF and Point-Spread
Function fitting routines in the {\tt daophot} package.  The {\it i'}
observations was conducted under photometric conditions, and the
photometric zeropoint was calibrated using observations of photometric
standards on the same night, assuming standard colour and extinction
terms.  The {\it g'} and {\it r'} observations were also conducted
under photometric conditions, however conditions degraded later in the
night and it was not possible to observe a photometric standard.
Using observations of photometric standards acquired in an interval
spanning 5 days either side of 2011 Sep 21, we have established that
the {\it g'} and {\it r'} photometric zeropoints are stable, with the error on
the weighted mean zeropoint of $0.005$ magnitudes.  The calibrated
{\it g'}, {\it r'} and {\it i' } photometry was originally in the AB
magnitude system and was converted to Vega magnitudes with the
addition of the constants +0.102, -0.165 and -0.403 mags, respectively,
derived through synthetic photometry of the spectrum of Vega
\citep{2004AJ....127.3508B}. 

\subsection{Late-time HST WFC3 UVIS \& IR imaging}
Additional late-time observations of the site of SN~2008bk were acquired with the $HST$ Wide Field Camera 3 Ultraviolet-Visible (UVIS) and Infrared (IR) Channels, with the $F814W$, $F125W$ and $F160W$ filters.  The observations were acquired as part of program GO-12262 (PI Maund).  For each filter, four separate exposures in a four-point box dither pattern.  The separate exposures were combined using the {\tt multidrizzle} package, running under {\sc PyRAF}, to achieve better sampling than the native pixel scale affords from a single pointing.  The observations were conducted using the smaller $1024\times 1024$ and $512 \times 512$ pixel subarrays of the UVIS2 and IR detectors.
The UVIS $F814W$ image was drizzled to a final pixel scale of $0.025\arcsec \, \mathrm{px^{-1}}$, while the IR observations were drizzled to a final pixel scale of $0.078\arcsec \,\mathrm{px^{-1}}$.  Photometry of the images was conducted using {\tt daophot}, running as part of {\sc iraf}.  Empirical aperture corrections were determined to final aperture sizes of $\mathrm 0.4\arcsec$, and the standard Vegamag zero points were adopted\footnote{http://www.stsci.edu/hst/wfc3/phot\_zp\_lbn}.  The UVIS2 $F814W$ image was further corrected for Charge Transfer Inefficiency using the calibration of \citet{wfc3cte}, following the scheme outlined by \citet{2008AJ....135.1900A} and \citet{2013arXiv1302.7152M}.   As there were no corresponding pre-explosion HST images, the late-time observations were used to probe the nature of the underlying background flux that may have contributed to the pre-explosion ground-based observations.

\subsection{Differential Astrometry}
The position of the SN in the pre-explosion and late-time images was determined using differential astrometry with respect to the position of the SN observed in a post-explosion VLT NACO adaptive-optics $K_{S}$image \citep[presented by][]{2008ApJ...688L..91M}.  Transformations between the post-explosion $K_{S}$ image and the late-time images were calculated using the {\sc iraf} task {\tt geomap}.  The post-explosion image was used to directly determine the position of the SN on the late-time $i'IYJHK_{S}$ images; and the positions on the $i'$ and $I$ images were further used to determine the SN position on the $g'r'$ and $BV$ images respectively.  The SN position on the pre-explosion frames was determined using transformations calculated between the pairs of late-time and pre-explosion images with the same filters. 

\subsection{Detection limits \& artificial star tests}
In each image, where objects of interest were not detected, the
detection threshold was determined using artificial star tests.
Artificial stars were placed in the images, at the positions where the
detection thresholds were to be determined, utilising the PSFs derived
from previous stages of photometry with {\tt daophot}. Recovery of the
artificial stars was then attempted using {\tt daophot} in the same
configuration used for the original photometry of the images.  The
completion function for the recovery of the artificial stars, as a
function of magnitude, was approximated by a cumulative normal
distribution, with mean at $50\%$ completeness and the standard
deviation corresponding to the breadth of the function.

\subsection{ISIS image subtraction}
Following \citet{2009Sci...324..486M}, we use the image subtraction package {\it ISIS} \citep{1998ApJ...503..325A,2000A&AS..144..363A} to subtract the late-time images from the pre-explosion images, to derive precise photometry of the now absent progenitor object. The late-times images were transformed to match the coordinate system of the pre-explosion images; as the late-time images were generally taken under better seeing conditions than the pre-explosion images, this resulted in a slight degradation of image quality of the late-time images, but still better than the pre-explosion images.  The late-time images were then used as the template frames to be subtracted from the pre-explosion images, except in the case of  ISAAC $J$ and $K_{S}$ frames for which the pre-explosion images were acquired under better seeing.  Aperture photometry was conducted on the residuals in the difference image, as part of the {\it ISIS} package (this was checked against aperture photometry with {\tt daophot}, and the differences were found to be negligible).  The zeropoint was determined relative to aperture photometry of non-varying stars in the same field and their photometry derived using {\tt daophot} (see section \ref{sec:obs}).  Although {\it ISIS} derives the usual statistical uncertainties from aperture photometry, the additional systematic uncertainty in using {\it ISIS}  was estimated by conducting multiple iterations of the image subtraction process with the settings modified.  This allowed us to quantify the effect of various parameters, such as: the number of stamps used to calculate the convolution kernel to match the PSFs of the template and input images; the degree of the fit to the background; the degree of the spatial variability of the kernel; and the size of the aperture and background annulus used for the photometry of the residuals. 

%%%%%%%%%%%%%%%%%%%%%%%%%%%%%%%%%%%%%%%%%%%%%%%%%%%%%%%%%%%%%%%%%%%%
%TABLE OBSERVATIONS - tab:obs
%%%%%%%%%%%%%%%%%%%%%%%%%%%%%%%%%%%%%%%%%%%%%%%%%%%%%%%%%%%%%%%%%%%%
\begin{table*}
\caption{\label{tab:obs} Pre-explosion, post-explosion and late-time observations of the site of SN~2008bk}

\begin{tabular}{lcccccc}
\hline \hline
Date                  &  Telescope+Instr.  & Filter   & Exposure     & Pixel                   & Seeing             & Airmass \\
(UT)                  &                    &          &$\mathrm{(s)}$& Scale $\mathrm{\arcsec}$& $\mathrm{\arcsec}$ &         \\ 
\hline
2007 Sep 05.25        &Gemini$-$S$+$GMOS   &$g^{\prime}$& $10\times200$&0.146                   &$0.71$                 &1.003    \\
2007 Sep 11.09        &Gemini$-$S$+$GMOS   &$r^{\prime}$& $15\times300$&0.146                   &$1.06$                 &1.319    \\
2007 Sep 05.29        &Gemini$-$S$+$GMOS   &$i^{\prime}$& $15\times200$&0.146                   &$0.72$                 &1.065    \\
\\
2011 Sep 21.21        &Gemini$-$S$+$GMOS   &$g^{\prime}$& $11\times200$&0.146                   &$0.58$                 &1.009    \\
2011 Sep 21.26        &Gemini$-$S$+$GMOS   &$r^{\prime}$& $10\times300$&0.146                   &$0.60$                 &1.055    \\
2011 Sep 24.15        &Gemini$-$S$+$GMOS   &$i^{\prime}$& $10\times300$&0.146                   &$0.41$                 &1.024    \\
\hline
2001 Sep 16.00       & VLT+FORS1  &$B_{BESS}$  & $300$ & 0.200 & 1.16 & 1.725  \\
2001 Sep 16.00       & VLT+FORS1  &$V_{BESS}$  & $300$ & 0.200 & 1.08 & 1.674 \\
2001 Sep 16.00       & VLT+FORS1  &$I_{BESS}$  & $480$ & 0.200 & 0.92 & 1.626 \\
\\
2011 Aug 18.24       & VLT+FORS   &$B_{BESS}$  &$3\times480$ &  0.252   &  0.58  &  1.036\\   
2011 Aug 21.26      & VLT+FORS   &$V_{BESS}$  &$3\times300$ &  0.252   &  0.73  &  1.017\\   
2011 Aug 18.23       & VLT+FORS   &$I_{BESS}$  &$5\times200$ &  0.252   &  0.55  &  1.061\\   
\hline 
2007 Oct 16.13         & VLT+HAWKI  &$Y$ &$6\times 60 $ &0.106   &   0.81  & 1.010    \\
2007 Oct 16.10       & VLT+HAWKI  &$H$ &$2\times60$ &0.106   & 0.77     & 1.021      \\
\\
2011 Sep 16.22       & VLT+HAWKI  &$Y$ &$5\times30$ & 0.106  &   0.34       &  1.015\\  
2011 Sep 16.30       & VLT+HAWKI  &$H$ &$5\times10$ & 0.106  &    0.36      &  1.153\\  
\hline 
2005 Oct 16.20      & VLT+ISAAC  &$J$     &$17\times60$& 0.148   & 0.50         & 1.352  \\  
2005 Oct 16.10       & VLT+ISAAC  &$K_{S}$ &$58\times60$& 0.148   & 0.37         & 1.112 \\
\\
2011 Sep 15.24       & VLT+ISAAC  &$J$     &$18\times30$& 0.148   & 0.53         &  1.018\\  
2011 Sep 15.24       & VLT+ISAAC  &$K_{S}$ &$18\times15$& 0.148   & 0.44         &  1.029\\
\hline
2008 May 19.35       & VLT+NACO                     &$K_{S}$      & $20 \times 69$ & $0.027$ &$0.1$    & 1.820 \\
\hline 
2011 Apr 29.99       & HST+UVIS2  &$F814W$ & 915        & 0.025   & $\cdots$ & $\cdots$ \\
2011 Apr 29.99       & HST+IR     &$F125W$ & 828.577    & 0.078   & $\cdots$ & $\cdots$ \\
2011 Apr 30.00       & HST+IR     &$F160W$ & 461.837    & 0.078   &$\cdots$ & $\cdots$ 
\\
\hline\hline
\end{tabular}

\end{table*}
%%%%%%%%%%%%%%%%%%%%%%%%%%%%%%%%%%%%%%%%%%%%%
%RESULTS
%RESULTS
%RESULTS
%RESULTS
%RESULTS
%RESULTS
%RESULTS
%RESULTS
%RESULTS
%RESULTS
%RESULTS
%RESULTS
%RESULTS
%RESULTS
%%%%%%%%%%%%%%%%%%%%%%%%%%%%%%%%%%%%%%%%%%%%%
\section{Results}
\label{sec:res}
The pre-explosion, late-time and difference images of the site of SN~2008bk are
presented for the VLT FORS, Gemini GMOS and VLT HAWK-I and ISAAC observations in Figs. \ref{fig:res:fors}, and \ref{fig:res:gem} and \ref{fig:res:ir}, respectively.  The late-time HST observations are presented in Fig. \ref{fig:res:hst}.  Photometry of the pre-explosion and late-time
images is presented in Table \ref{tab:res:phot}, while photometry of the late-time HST observations is presented in Table \ref{tab:res:hstphot}.  In comparing the pre-explosion and late-time observations (and the corresponding difference images), particularly in the IR, the object detected in the pre-explosion images is clearly no longer present; confirming the original identification of this star as the progenitor of SN~2008bk.
%%%%%%%%%%%%%%%%%%%%%
%RESULTS BVI
%%%%%%%%%%%%%%%%%%%%%
\subsection{VLT FORS B,V \& I}
\label{sec:res:BVI}
Our reanalysis of the pre-explosion {\it FORS} {\it BV} frames, in which
the SN position was identified with a precision of $0.077\arcsec$, confirms the conclusion of \citet{2008ApJ...688L..91M}: no single point source was present at the SN position.  The new brightness limits determined for the
pre-explosion {\it BV} images, having been calibrated against the corresponding photometry of the late-time images, are brighter than derived by
\citeauthor{2008ApJ...688L..91M} (which were inconsistent
the brighter $\sim V$ magnitude calculated by
\citet{2012AJ....143...19V} from the pre-explosion GMOS images). Our
brightness limits, calculated using artificial star tests, take into
account nearby sources in trying to recover a single point source at
the SN position and, hence, we believe the brightness limits derived
in this fashion are more reliable. 

\begin{figure*}
\includegraphics[width=15cm]{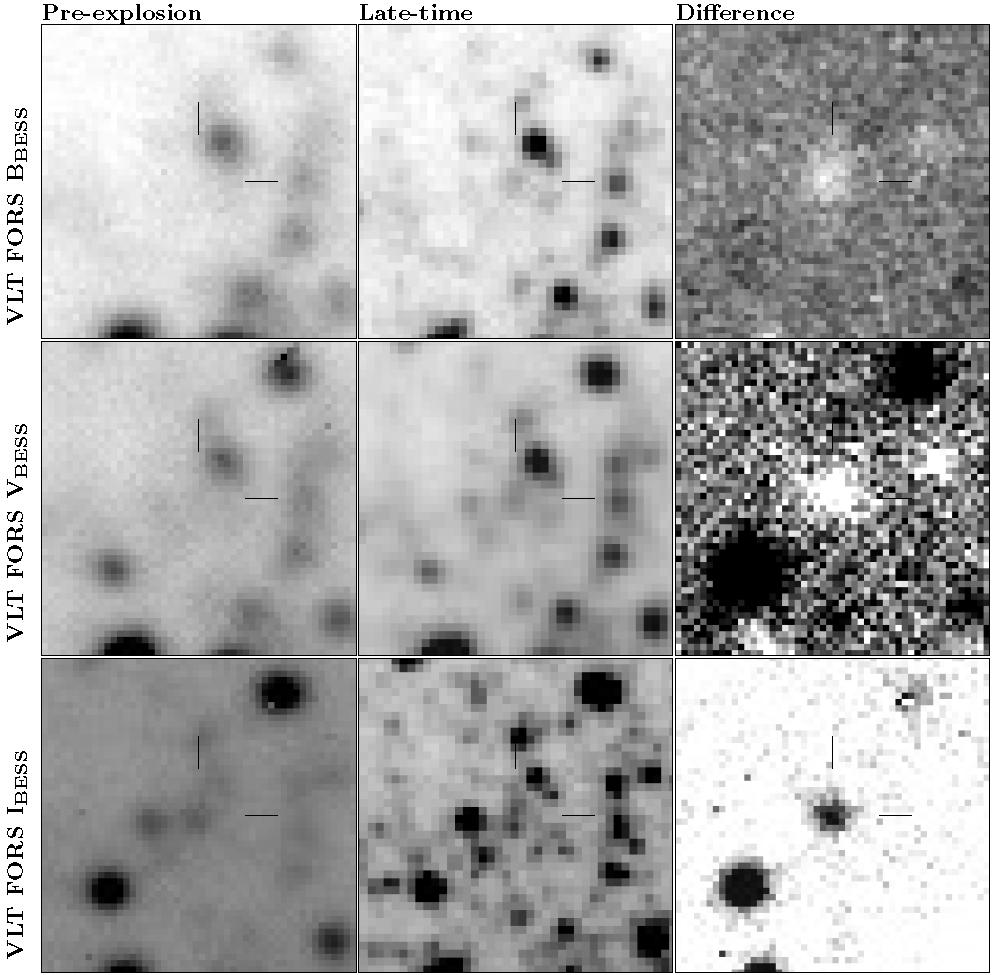}
\caption{Pre-explosion and late-time VLT FORS images of the site of SN~2008bk, in the $B$, $V$ and $I$ filters.  The difference image, between the pre-explosion and late-time frames, is shown in the righthand column.  The images have dimensions $10\arcsec \times 10\arcsec$ and are oriented such that North is up, East is left.   The position of SN~2008bk, in all the images, is indicated by the cross hairs.  In the difference images positive (lighter) residuals correspond to source brighter in the late-time images, while negative (darker) residuals correspond to sources that are fainter in the late-time images.   Details of these observations are presented in Table \ref{tab:obs}.}
\label{fig:res:fors}
\end{figure*}

The
bright positive residual in the {\it FORS} {\it BV} difference images (see Fig. \ref{fig:res:fors}), corresponding to an increase in brightness between the pre-explosion and late-time observation, appears much
more diffuse than nearby point source residuals.  The photometry of the source in the late-time $BV$ images yields sharpness values of $0.7$ and $1.1$ respectively, implying that the SN residual has a broader and flatter profile than point sources.  

The position of the SN on the pre-explosion {\it FORS} {\it I}-band image was identified to within $0.05\arcsec$, and we identify the same source found by \citet{2008ApJ...688L..91M} as the progenitor.  The PSF-fitting photometry yields values similar to those found by \citeauthor{2008ApJ...688L..91M}, with $\chi^{2}=0.8$ and sharpness of 0.2 (consistent with a point source).  Inspection of the late-time images clearly shows a large flux deficit at the SN position in the $I$-band image, although there is still a faint source coincident with the SN position (estimated to within $0.03\arcsec$).  As for the pre-explosion images, the PSF-fitting photometry yields values consistent with a point source ($\chi^{2}=1.4$, sharpness$=-0.1$), unlike the residuals in the late-time $BV$ images.  \citeauthor{2008ApJ...688L..91M} observed a bright source to the South of the SN position in the pre-explosion $ISAAC$ $K_{S}$-band images, and we find a source at the same offset and position angle in the late-time $I$-band image (see section \ref{sec:res:hst}).  The flux deficit between the pre-explosion and late-time $I$-band observations is clearly evident in the difference image (Fig. \ref{fig:res:fors}), and analysis of the residual is consistent with a point source.  The source recovered in the late-time ground-based $I$-band image was found, in the corresponding HST observations (see section \ref{sec:res:hst}), to have a contributions from the fading SN; requiring an additional correction to the photometry derived from the difference images.  The derived brightness for the progenitor (including the correction for late-time SN flux) is significantly lower that the brightness determined from the pre-explosion images alone, but similar to the $I$-band brightness measured by \citeauthor{2008ApJ...688L..91M}.
%%%%%%%%%%%%%%%%%%%%%
%RESULTS gri
%%%%%%%%%%%%%%%%%%%%%
\subsection{Gemini GMOS-S {\it g'}, {\it r'} \& {\it i'}}
\label{sec:res:gri}
\begin{figure*}
\includegraphics[width=15cm]{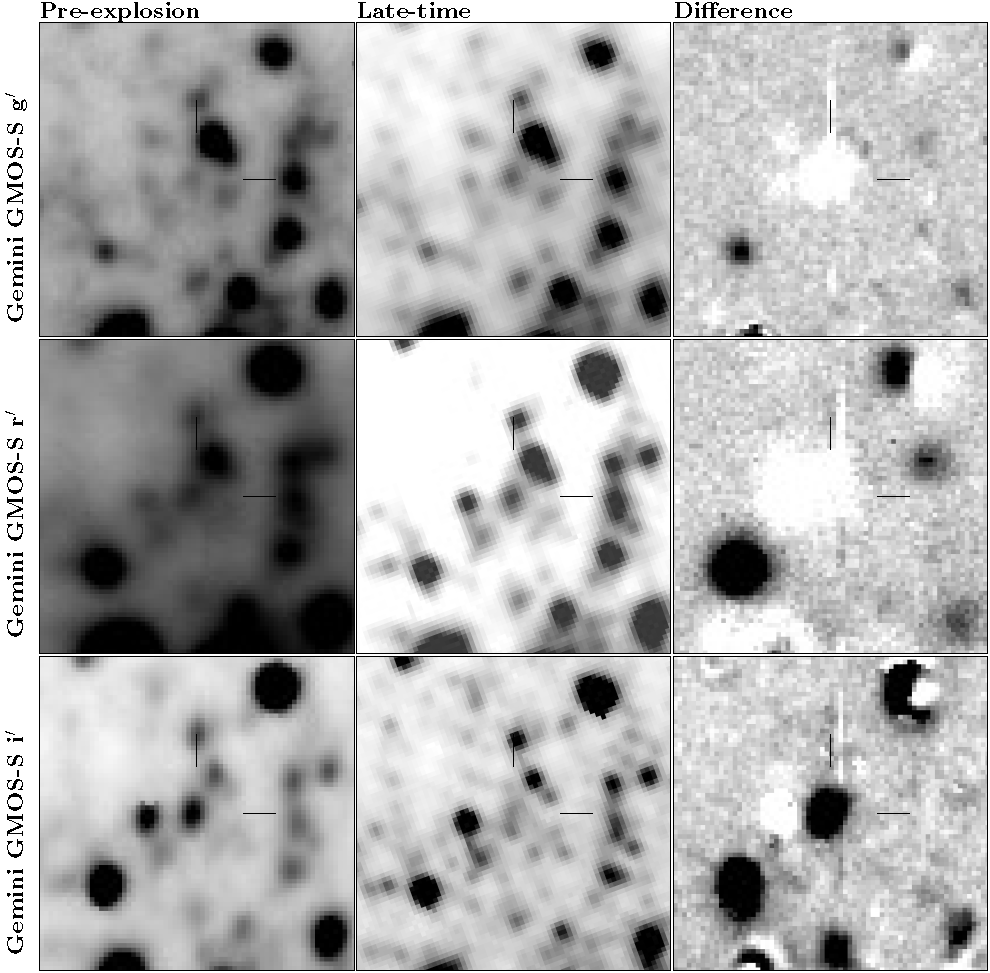}
\caption{Pre-explosion and late-time Gemini GMOS-S imaging of the site of SN~2008bk, in the $g^{\prime}$, $r^{\prime}$ and $i^{\prime}$ filters.  For a description of the columns, see Figure \ref{fig:res:fors}.}
\label{fig:res:gem}
\end{figure*}

\citet{2012AJ....143...19V} reported the detection of a point source at
the SN position in pre-explosion Gemini $g'r'$ images.  In
conjunction with the post-explosion NACO $K_{S}$ image, we identify
the same source in the pre-explosion GMOS $g'r'$ images, with an
uncertainty of $0.05\arcsec$.  Fitting the PSF to the pre-explosion
source yielded $\chi^{2} \sim 1$ while the sharpness parameter values,
0.50 and 0.68 for $g'$ and $r'$ respectively, are suggestive that
the object is slightly extended; possibly due to the extended wings from the source to the North-West of the SN position (see Fig. \ref{fig:res:gem}).  The photometry of this
source in the GMOS images is approximately consistent with the upper
limits placed on the source brightness in the pre-explosion {\it VLT}
{\it FORS} images.  Significant flux is still recovered at the SN position in the late-time $g'$ and $r'$ images, that is both brighter and more extended than the pre-explosion source and, similarly to the late-time  {\it FORS} {\it BV} images, has a flatter, more extended profile.  As the SN is
still significantly brighter than the progenitor candidate (or
detection limits thereon), image subtraction techniques cannot be used
to probe the nature of the progenitor itself.

It is clear from
Figs. \ref{fig:res:fors} and \ref{fig:res:gem}, that there is
additional extended flux around the SN position in the late-time
images that was not present in the pre-explosion images.  In the
late-time Gemini GMOS $g'r'$ images, the apparent extended nature of the
residual is clearer, especially in the $r'$-band difference image
(where the effect of \halpha emission, either from the recombination
in the local environment or from a light echo, is expected to be
large - see section \ref{sec:res:hst}).  The shapes of the pre-explosion source in the $g^{\prime}r^{\prime}$ images may suggest that crowding with nearby objects may cause the measured brightness of the progenitor in these two filters to be, possibly, overestimated (see below).

The SN position was identified to within $0.04\arcsec$ and $0.03\arcsec$ on the pre-explosion and late-time \sloani images, respectively.  In both pre-explosion and late-time \sloani images, we recover a source at the SN position, whose properties are consistent with a point source.  In the corresponding difference image, we see a negative residual due to the large flux deficit at the SN position between the two epochs.  We also observe that the object immediately to the East of the SN progenitor has increased in brightness. 
As for the $I$-band observation, we apply a small correction to the \sloani brightness derived from the difference image, due to residual SN flux in the late-time image, derived from the late-time HST observations (see section \ref{sec:res:hst}).

We find the $i^{\prime}$ brightness of the progenitor, derived using image subtraction techniques, to be significantly less than determined by direct photometry on the pre-explosion images alone.  This suggests that, although the pre-explosion \sloani photometry is dominated by the progenitor flux, it is blended with contributions from surrounding stars (as was also observed for the $I$-band observations).  This may imply that a similar effect may apply to the photometry of the pre-explosion \sloang and \sloanr observations.  While the \sloani photometry is corrected for any contribution from blended flux, through template subtraction, the \sloang and \sloanr observations are not and may, therefore, lead to an overestimate of the progenitor brightness.  The additional uncertainties associated with the photometric zeropoints for the late-time \sloang and \sloanr observations suggests that these two photometric measurements of the progenitor should be treated with some caution.
%%%%%%%%%%%%%%%%%%%%%
%RESULTS YJHK
%%%%%%%%%%%%%%%%%%%%%
\subsection{VLT HAWK-I/ISAAC Y, J, H \& $K_{S}$}
\label{sec:res:yjhk}
The VLT ISAAC $J$ and $K_{S}$ and HAWKI $H$ images were common to the studies presented by \citet{2008ApJ...688L..91M} and \citet{2012AJ....143...19V}.  A pre-explosion $Y$-band observation was available in the archive, however it could not be photometrically calibrated in the original studies. The SN position was identified on the late-time $YJHK_{S}$ frames to within $0.016$, $0.052$, $0.024$ and $0.037\arcsec$, respectively.  A bright source is detected in all the pre-explosion $YJHK_{S}$ images at the SN position.  The source appears consistent with a single point source in the $JHK_{S}$ frames.  PSF fitting to the source recovered in  the $Y$ frame yields  $\chi^{2}=3.8$, suggesting it is possibly an extended source; however this may also be due to the poorer quality of the data compared to the $JHK_{S}$ images.   A source is subsequently recovered at the SN position in the late-time $Y$ band image, although significantly fainter than the pre-explosion source.   The SN is not detected as a point source in the late-time $JHK_{S}$ images, to levels significantly fainter than the pre-explosion source; however, we note that there is diffuse emission at and around the SN position (and this is discussed in Section \ref{sec:res:hst}).   \citet{2008ApJ...688L..91M} measured the PSF of the pre-explosion $K_{S}$ source as being extended, hypothesising the pre-explosion source was a blend with a source located $\sim 0.5\arcsec$ South of the SN position.  This source is recovered in the late-time $K_{S}$ images.
\\
The differences between the pre-explosion and late-time near-infrared images show significant residuals at the SN position consistent with the disappearance of the source detected in the pre-explosion images.  In all cases, the corresponding residual is consistent with a point source.  For the $Y$ and $J$ band observations, we find that the progenitor brightness determined using image subtraction techniques is fainter than estimated from the pre-explosion images alone.   As for the $I$ and \sloani observations, the difference suggests that blended flux from other sources contaminates the pre-explosion source.  A small correction derived from the late-time HST observations was used to correct  the progenitor's brightness determined from the $Y$ and $J$ difference images for possible residual SN flux.  The $K_{s}$ brightness derived from the difference image is slightly fainter than measured from the pre-explosion images, and may suggest contamination from the source to the South was not properly accounted for in photometry of the pre-explosion frame.  The $H$-band photometry of the progenitor is identical for photometry of the pre-explosion images and from the corresponding difference image.  In the $ISAAC$ difference images ($J$ and $K_{s}$), additional residuals are also seen for some nearby stars.  The peculiar nature of these residuals, compared to the progenitor, suggests a possible deficiency with the calculated convolution kernel used to match the PSFs of the pre-explosion and late-time images.  We note, however, that the scale of these residuals is small and that photometry of these residuals averages to zero brightness.
%%%%%%%%%%%%%%%%%%%%%
%RESULTS HST
%%%%%%%%%%%%%%%%%%%%%
\subsection{HST WFC3 $F814W$, $F125W$ \& $F160W$}
\label{sec:res:hst}
The late-time HST $F814W$, $F125W$ and $F160W$ observations of the site of SN~2008bk are presented on Figure \ref{fig:res:hst}.  Photometry of the sources recovered at and around the SN position is reported in Table \ref{tab:res:hstphot}.  In addition photometry was also conducted on late-time WFC3 UVIS $F336W$, $F555W$ and $F814W$ observations acquired as part of program GO-12285 (PI Soria) on 2011 May 07.  Photometry of these images was conducted using the {\sc dolphot} package\footnote{http://americano.dolphinsim.com/dolphot/} \citep{dolphhstphot}.

\begin{figure*}
\includegraphics[width=15cm]{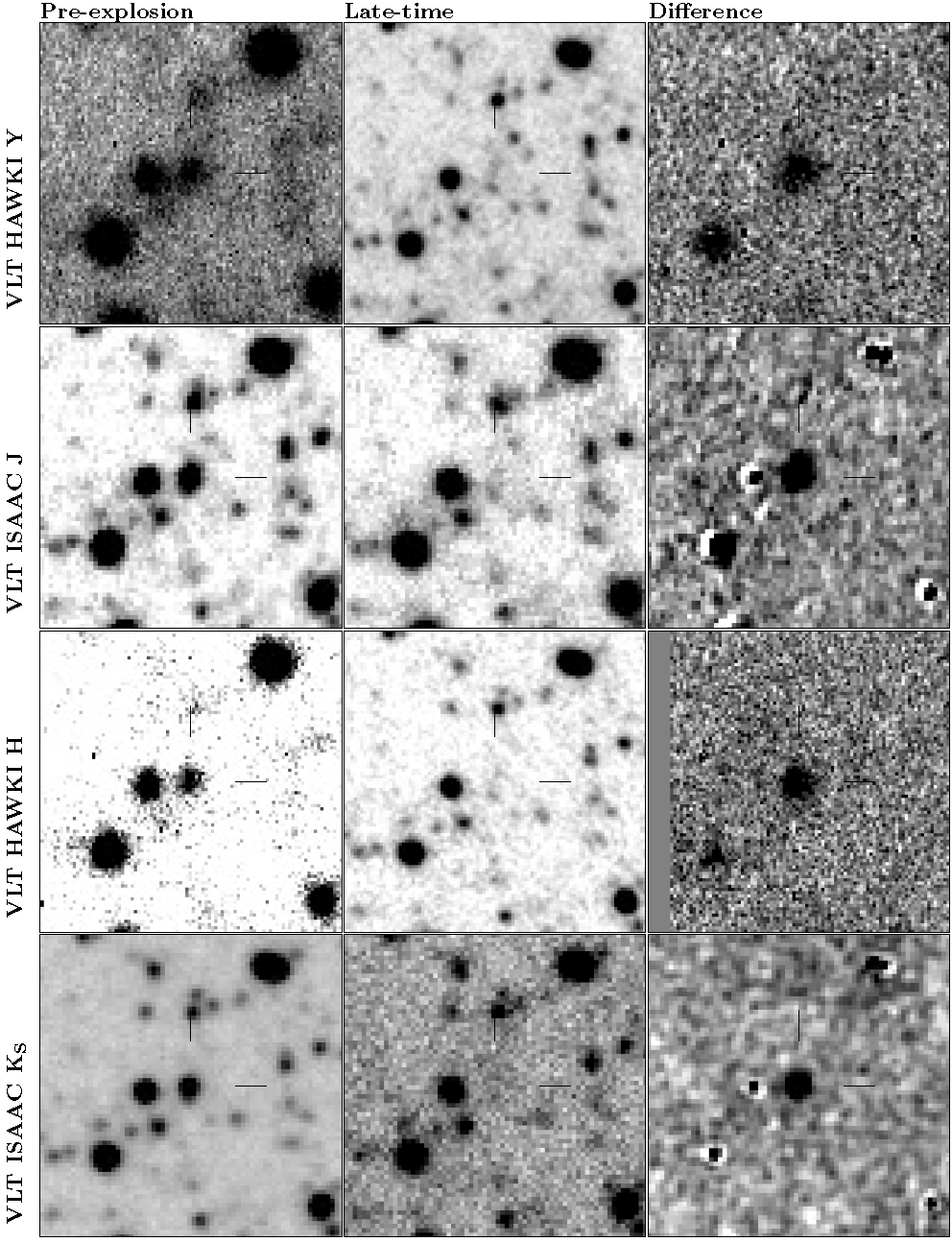}
\caption{Pre-explosion and late-time VLT Near-Infrared HAWK-I and ISAAC imaging of the site of SN~2008bk, in the $Y$, $J$, $H$ and $K_{S}$ filters. For a description of the columns, see Figure \ref{fig:res:fors}.}
\label{fig:res:ir}
\end{figure*}

\begin{figure*}
\includegraphics[width=15cm]{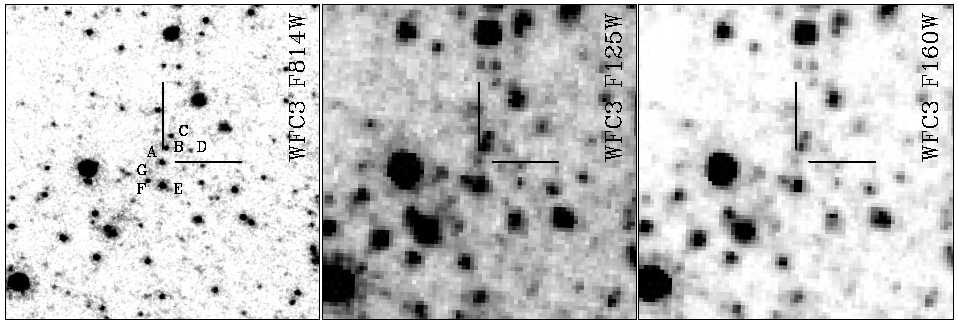}
\caption{Late-time HST WFC3 UVIS and IR channel imaging of the site of SN~2008bk, with the $F814W$, $F125W$ and $F160W$ filters.  The images are centred of the SN position, indicated by the cross hairs (labeled A), and oriented such that North is up, East is left.  Each of the frames has dimensions $6\arcsec \times 6\arcsec$.  The SN and surrounding stars are labeled $A-G$ (see text).}  
\label{fig:res:hst}
\end{figure*}

The SN is still recovered in all the images and, as reported by \citet{2013arXiv1305.6639V}, there is evidence for a light echo around the SN position.  There are a number of sources around the SN (labeled in Figure \ref{fig:res:hst}) which are bright in the IR.  The brightest nearby IR source is labeled Source E, which is the same source identified to the South of the SN position in the pre-explosion and late-time images.  The diffuse emission recovered in the late-time ground based IR images (see Fig. \ref{fig:res:ir} and Section \ref{sec:res:yjhk}) is shown to be a blend of SN emission,  contributions from nearby stars and possibly the light echo.  At redder wavelengths, the  "diffuse emission" observed in the ground-based images clearly arises from the surrounding stars.  Only one source in the vicinity of SN~2008bk is observed to have a blue SED (source G), however it is not as bright as the pre-explosion source recovered at the SN position in the ground-based pre-explosion optical images.  We note, however, that the background at the SN position in the $F336W$ and $F555W$ is uneven and the wings of the bright blue sources to the North-West may have contributed to the flux measured in the pre-explosion Gemini GMOS-S $g^{\prime}$ and $r^{\prime}$ images.  Although there are a number of IR sources around the SN position, these are significantly fainter than pre-explosion photometry of the progenitor (see Table \ref{tab:res:hstphot}).  Since the IR brightness of the progenitor is derived from the difference images, these surrounding stars do not contaminate the progenitor photometry when derived in this manner.

As the SN is still recovered in the late-time HST observations, the photometry recovered from ground based images with late-time image subtraction must be corrected for the small amount of residual flux remaining at the SN position.  We evaluated the corrections from the $F814W$ and $F125W$ images to be $\Delta I\approx \Delta i^{\prime}=-0.1$ and $\Delta J = -0.025$; although we do not have a late-time HST observations corresponding to the ground-based $Y$ filter, given the colour of the SN and the progenitor we interpolate between the corrections and estimate $\Delta Y \approx -0.05$ -- $-0.08$.  We note that these corrections are consistent with the differences between the pre-explosion and difference image photometry.  
%%%%%%%%%%%%%%%%%%%%%%%%%%%%%%%%%%%%%%%%%%%%%%%%%%%%%%%%%%%%%%%%%%%%
%FIGURE VLT FORS - fig:res:fors
%%%%%%%%%%%%%%%%%%%%%%%%%%%%%%%%%%%%%%%%%%%%%%%%%%%%%%%%%%%%%%%%%%%%
%%%%%%%%%%%%%%%%%%%%%%%%%%%%%%%%%%%%%%%%%%%%%%%%%%%%%%%%%%%%%%%%%%%%
%FIGURE GEMINI GMOS - fig:res:gem
%%%%%%%%%%%%%%%%%%%%%%%%%%%%%%%%%%%%%%%%%%%%%%%%%%%%%%%%%%%%%%%%%%%%
%%%%%%%%%%%%%%%%%%%%%%%%%%%%%%%%%%%%%%%%%%%%%%%%%%%%%%%%%%%%%%%%%%%%
%FIGURE VLT HAWKI ISAAC - fig:res:ir
%%%%%%%%%%%%%%%%%%%%%%%%%%%%%%%%%%%%%%%%%%%%%%%%%%%%%%%%%%%%%%%%%%%%
%%%%%%%%%%%%%%%%%%%%%%%%%%%%%%%%%%%%%%%%%%%%%%%%%%%%%%%%%%%%%%%%%%%%
%FIGURE HST IMAGING - fig:res:hst
%%%%%%%%%%%%%%%%%%%%%%%%%%%%%%%%%%%%%%%%%%%%%%%%%%%%%%%%%%%%%%%%%%%%
%%%%%%%%%%%%%%%%%%%%%%%%%%%%%%%%%%%%%%%%%%%%%%%%%%%%%%%%%%%%%%%%%%%%
%TABLE PROGENITOR PHOTOMETRY - tab:res:phot
%%%%%%%%%%%%%%%%%%%%%%%%%%%%%%%%%%%%%%%%%%%%%%%%%%%%%%%%%%%%%%%%%%%%
\begin{table*}
\caption{\label{tab:res:phot}Pre-explosion, late-time and difference image photometry of the site of SN~2008bk}
\begin{tabular}{lccc|c|cc}
\hline\hline
Band        & Pre-explosion & Late-time   & Difference & Progenitor     & Mattila et al. & Van Dyk et al.\\
            &               &             &            &                &\\
\hline
{\it B}     & $>22.77(0.17)$&$23.12(0.13)$& $\cdots$   &  $>22.77(0.17)$& $>22.9$        &\\
{\it V}     & $>22.19(0.05)$&$21.97(0.08)$& $\cdots$   &  $>22.19(0.05)$&  $>23.0$       &$22.81(0.09)$\\
{\it I}     & $20.81(0.04)$ &$22.99(0.14)$&21.36(0.05) & $21.26(0.06)^{\dagger}$     & $21.20(0.19)$  &$20.71(0.08)$\\
   & && & &       &   \\
{\it g'}    & $23.96(0.10)$ &$22.65(0.04)$& $\cdots$   &  $23.56(0.04)$ & $\cdots$       & $\cdots$ \\
{\it r'}    & $22.47(0.07)$ &$22.34(0.05)$& $\cdots$   &  $22.27(0.02)$ & $\cdots$       & $22.03(0.07)^{\ddagger}$ \\
{\it i'}    & $20.89(0.01)$ &$23.06(0.07)$&21.42(0.06) & $21.32(0.06)^{\dagger}$& $\cdots$       & $\cdots$  \\
   & && & &       &   \\
{\it Y}     & $19.88(0.11)$ &$22.69(0.11) $&20.13(0.18)&$20.05(0.17)^{\dagger}$    & $\cdots$& $\cdots$\\
{\it J}     & $19.34(0.02)$ &$>22.67(0.10)$&19.47(0.03)&$19.47(0.03)$   & $19.50(0.06)$& $19.26(0.14)$\\
{\it H}     & $18.54(0.04)$ &$>20.08(0.06)$&18.52(0.04)&18.52(0.04)     & $18.78(0.11)$&$18.55(0.06)$\\
{\it $K_{S}$}& $18.31(0.03)$ &$>20.13(0.12)$&18.39(0.03)&$18.39(0.03)$  & $18.34(0.07)$&$18.14(0.10)$\\
\hline\hline
\end{tabular}
\\
$^{\dagger}\,$ Progenitor magnitude is a combination of the pre-explosion/late-time brightness difference and the late-time brightness.\\
$^{\ddagger}\,$ $R$-band photometric measurements derived from an underlying $r^{\prime}$ observation.
\end{table*}

%%%%%%%%%%%%%%%%%%%%%%%%%%%%%%%%%%%%%%%%%%%%%%%%%%%%%%%%%%%%%%%%%%%%
%TABLE LATE-TIME HST PHOT - tab:res:hstphot
%%%%%%%%%%%%%%%%%%%%%%%%%%%%%%%%%%%%%%%%%%%%%%%%%%%%%%%%%%%%%%%%%%%%
\begin{table*}
\caption{\label{tab:res:hstphot} Late-time HST photometry of SN~2008bk and surrounding stars}
\begin{tabular}{cccccccc}
\hline\hline
       & \multicolumn{3}{c}{2011 May 07}    &        &\multicolumn{3}{c}{2011 Apr 29}  \\
 \cline{2-4}\cline{6-8}\\[-0.75ex]
       &   $F336W$    &  $F555W$     & $F814W$   &   &    $F814W$   &   $F125W$    & $F160W$  \\
\hline
A(SN)      &$23.57\pm0.05$&$24.14\pm0.02$&$24.01\pm0.04$&&$23.84\pm0.05$&$24.50\pm0.07$&$23.95\pm0.23$\\
B      &  $\cdots$    &$26.36\pm0.08$&$24.16\pm0.04$&&$24.06\pm0.05$&$22.63\pm0.05$&$22.19\pm0.06$\\
C      &  $\cdots$    &$26.49\pm0.09$&$24.28\pm0.04$&&$24.21\pm0.05$&$22.85\pm0.06$&$22.55\pm0.06$\\
D      &  $\cdots$    &$26.57\pm0.10$&$24.89\pm0.07$&&$24.86\pm0.06$&$23.74\pm0.08$&$23.16\pm0.08$\\
E      &  $\cdots$    &$25.28\pm0.04$&$23.10\pm0.02$&&$23.00\pm0.04$&$21.82\pm0.06$&$21.48\pm0.13$\\
F      &  $\cdots$    &$26.01\pm0.07$&$24.24\pm0.04$&&$24.22\pm0.06$&$22.84\pm0.08$&$22.31\pm0.12$\\
G      &$24.20\pm0.07$&$25.35\pm0.04$&$25.50\pm0.17$&&$25.36\pm0.10$&$\cdots$      &$\cdots$ \\
\hline\hline
\end{tabular}
\end{table*}
%%%%%%%%%%%%%%%%%%%%%%%%%%%%%%%%%%%%%%%%
%ANALYSIS
%ANALYSIS
%ANALYSIS
%ANALYSIS
%ANALYSIS
%ANALYSIS
%ANALYSIS
%ANALYSIS
%ANALYSIS
%ANALYSIS
%ANALYSIS
%ANALYSIS
%ANALYSIS
%ANALYSIS
%%%%%%%%%%%%%%%%%%%%%%%%%%%%%%%%%%%%%%%%
\section{Analysis}
\label{sec:ana}
%%%%%%%%%%%%%%%%%%%%%%%%%%%%%%%%%%%%%%%%
%DISTANCE
%%%%%%%%%%%%%%%%%%%%%%%%%%%%%%%%%%%%%%%%
\subsection{Distance, Metallicity \& Reddening}
\label{sec:ana:dist}

Since the explosion of SN~2008bk, there have been two new recent
distance measurements for the host galaxy determined using
the tip of the red giant branch (T-RGB; \citealt{2009AJ....138..332J})
and Cepheids \citep{2010AJ....140.1475P}, yielding $\mu=27.79\pm0.08$
and $27.68\pm0.09$ mag, respectively.  Given the agreement between the
two most recent measurements, we adopt the mean for this study
$\mu=27.74\pm0.06$, corresponding to $3.5\pm0.1\,\mathrm{Mpc}$.  The
analysis of the progenitor of SN~2008bk presented by \citet{2008ApJ...688L..91M}
utilised a larger distance $\mu=27.96\pm0.24$
\citep{2003A&A...404...93K}, also derived from the TRGB, while \citet{2012AJ....143...19V} adopted the recent Cepheid
distance above (although just using the systematic uncertainty); this
gives rise to differences in distance modulus of $\Delta \mu = -0.22$
and $\Delta \mu = 0.06$ between our study and those of
\citet{2008ApJ...688L..91M} and \citet{2012AJ....143...19V},
respectively.

We recalculated the deprojected offset of the SN from the centre of the host galaxy, as previously determined in \citet{2008ApJ...688L..91M}.  We find a smaller distance $r/R_{25}=0.24$.  This has consequences for the estimated metallicity at the SN position.  The analysis of \citet{2008ApJ...688L..91M} assumed an LMC metallicity.  Using the relations of \citet{metapil}, we derive an oxygen abundance at the radius of SN~2008bk of $12+\log \left(\frac{O}{H}\right)=8.42\pm0.07$.  Assuming solar and LMC metallicities corresponding to abundances of 8.65 and 8.35 \citep{2007A&A...466..277H}, this corresponds to $\log (Z/Z_{\odot}) = -0.23$, just higher than the LMC value.  Using the abundances derived to 27 H{\sc ii} regions, \citet{2010MNRAS.405.2737B} derived a similar oxygen abundance-radius relation using the \citet{2004MNRAS.348L..59P} $O3N2$ indicator, yielding a corresponding metallicity at the radius of SN~2008bk of 8.52 (with average uncertainty on the sample of measurements of $\pm 0.1 \mathrm{dex}$).    \citet{2012AJ....143...19V} measured the oxygen abundance for two H {\sc ii} regions, offset from the SN position, to be 8.45 and 8.52.  These values are suggestive that the metallicity appropriate for the progenitor of SN~2008bk is likely to be intermediate between the LMC and solar values.

The Galactic foreground reddening towards SN~2008bk is $E(B-V)=0.017$ mag \citep{2011ApJ...737..103S}.  \citet{2010MNRAS.405.2737B} measured the Balmer decrement for 29 {\sc H ii} regions in NGC 7793, finding $E(B-V)=0.18\pm0.02$ mags.  \citet{2008ApJ...688L..91M} used the line strengths reported by \citet{1985ApJS...57....1M} to determine a reddening to the {\sc H ii} region W13, located $1.5\arcmin$ from SN~2008bk, to be 0.45 mags.  The different levels of reddening measured towards {\sc H ii} regions suggest the possible role of significant and quite variable internal extinction inside NGC 7793.  Reddening estimates derived from SN 2008bk are also conflicted.  \citet{2012AJ....143...19V} suggested the absence of Na {\sc i} D in the early photospheric spectrum of SN~2008bk, and its photometric similarity to SN~1999br, are indicative of negligible host reddening; whereas \citet{2008CBET.1335....1M} claimed SN 2008bk was similar SN~1999em, which was subject to $E(B-V)\sim 0.1$ \citep{2000ApJ...545..444B,2008ApJ...688L..91M}.

\subsection{Luminosity constraints for the progenitor from the $K$-band brightness}
\label{sec:ana:klum}
\begin{figure} 
\includegraphics[width=7cm]{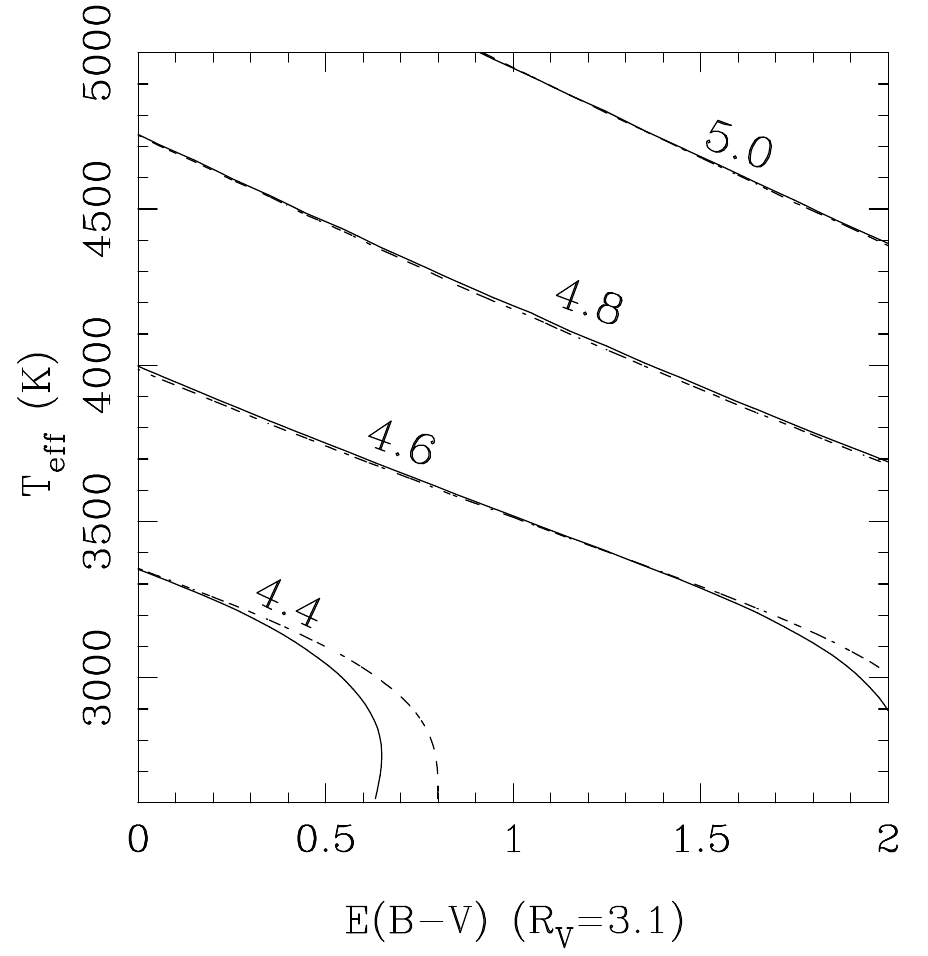}
\caption{Constraints on the luminosity of the progenitor of SN~2008bk from the observed $K$-band brightness (under the assumption of an $R_{V}=3.1$ reddening law).  The solid and dashed contours corresponds metallicities of $\log (Z/Z_{\odot})=-0.25$ and $0.0$, respectively.}
\label{fig:ana:klum}
\end{figure}

For standard values of $R_{V}$ (e.g. 3.1), the effects of extinction are significantly reduced in the $K$-band compared to optical bands.  The $K$-band magnitude derived for the progenitor of SN~2008bk, therefore, provides a reasonable constraint on the likely range of the luminosity of the progenitor.  For MARCS model SEDs (the $5M_{\odot}$ spherical models with $\log g=0.0$; \citealt{marcsref}),  we considered the temperature range $2600-5000K$ (at approximately LMC and solar metallicities).
Over this temperature range  the bolometric correction with respect to $K$ only changes by $0.8$ mags, while $BC_{V}$ changes by $\sim 5.5$ mags.  While the interpretation of the $K$-band magnitude does depend on the extinction and temperature, the effects of these parameters are relatively small.  On Figure \ref{fig:ana:klum} we show luminosity contours, over the full temperature range and reddenings in the range $0 \leq E(B-V) \leq 2$, given the $K$-band brightness measured for the progenitor from the difference imaging.  From the $K$-band brightness alone, the luminosity of the progenitor is constrained to lie in the range $4.3 \leq \log \left( L/L_{\odot}\right) \leq 5.1$.  Using the end point luminosities from STARS stellar evolution models for an LMC metallicity \citep{2008arXiv0809.0403S}, this limits the progenitor to having an initial mass $<15-17M_{\odot}$.

%%%%%%%%%%%%%%%%%%%%%%%%%%%%%%%%%%%%%%%%%%%%%%%%%%%%%%%%%%%%%%%%%%%%
%FIGURE K-BAND LIMITS - fig:ana:klum
%%%%%%%%%%%%%%%%%%%%%%%%%%%%%%%%%%%%%%%%%%%%%%%%%%%%%%%%%%%%%%%%%%%%
%%%%%%%%%%%%%%%%%%%%%%%%%%%%%%%%%%%%%%%%%%%%%%%%%%%%%%%%%%%%%%%%%%%%
\subsection{SED fitting}
\label{sec:ana:sed}
We considered the observed photometry, and upper limits, with respect
to SEDs derived from MARCS synthetic spectra \citep{marcsref}.  The
SEDs fits were conducted using our own Bayesian SED fitter (Maund, 2013, in prep),
which allow us to probe the effects of different metallicities and
reddening laws on the interpretation of the observations. 

We
considered the $5M_{\odot}$ spherical MARCS models with surface gravity $\log g = 0$ and
metallicities equivalent to $\log (Z/Z_{\odot})= -0.25$ ("LMC") and solar metallicity $(\log (Z/Z_{\odot})= 0.0)$.  In order to probe different types of
dust in the ISM in the host galaxy we computed synthetic photometry
for reddening laws with $R_{V}=2,3.1,4$ and $5$ \citep{ccm89}. We also considered synthetic photometry for a mixture of
dust components, with foreground Galactic reddening $0 \leq E(B-V)
\leq 0.2$ (consistent with the reddening estimated by \citet{2012AJ....143...19V}) and a Silicate dust component with $ 0 \leq \tau \leq 20$
\citep{2012ApJ...759...20K}.\\

To assess the sensitivity of the fits to particular data
points (specifically the uncertain $g'$ and $Y$ photometry) we
considered four groups of the photometric measurements containing all
or subsets of the data.  Each dataset is composed $N$ photometric
measurements and $M$ upper limits.  Our Bayesian SED fitter, based on the Nested Sampling algorithm \citep{2004AIPC..735..395S}, calculates
the Bayesian evidence (the marginal likelihood) as a measure of the relative quality of the fits to different models to
the same dataset.  For each dataset, we consider the relative Bayesian
evidence for each model (i.e. the Bayes factor) with respect to the
model with LMC metallicity and $R_{V}=3.1$.  We report Bayes factors
($K$; in decibans) and, in this instance, $\chi^{2}$ values. In the general case of considering datasets containing upper limits, the meaning of
$\chi^{2}$ and the corresponding $p$-values depart from their normal
definitions.  The upper limits provided by the pre-explosion FORS $BV$ photometry are not constraining and do not, in this case, impact the definition of the $\chi^{2}$ $p$-value.   The outcome of the SED fits are reported in Table
\ref{tab:ana:sedfit}.

Before considering the problem of parameter estimation, it is clear
from Table \ref{tab:ana:sedfit} that, in general, the solar
metallicity models provide a slightly worse fit to the observational data
(regardless of the choice of the underlying dataset) but not at a greatly
significant level.  Table \ref{tab:ana:sedfit} also highlights the
role of the $g'$ photometric point.  Those datasets containing the
$g'$ measurement provide a significantly worse $\chi^{2}$ than datasets without
and also favour reddening laws with higher $R_{V}$.  We note that
large values for $\chi^{2}$ achieved for datasets containing
$g^{\prime}$ are principally due to the MARCS models under-predicting
the flux, relative to the observed value, at that wavelength.
Conversely, the inclusion of the $Y$-band photometric point has little
impact on the quality of resulting fits.  For datasets in which the
$g^{\prime}$ point is not included, we see that there is little
evidence for achieving better fits with higher $R_{V}$ (although the
low value of $R_{V}=2$ provides a generally worse fit).  The dust mixture model
(with ISM and Silicate dust composition) is also disfavoured.  

The observed photometry could also be compared with MARCS models with different stellar mass, in particular the spherical $15M_{\odot}$ models.  At the time of writing these models are only available for $\log g = 0.0$ at solar metallicity, and cover a smaller temperature range ($3300-4500K$).  We have compared the synthetic photometry of the $5$ and $15M_{\odot}$ solar metallicity models and find that the differences in predicted colours predominantly affect bluer colours and are generally small and reddening independent.  The largest difference is for the $g^{\prime}-r^{\prime}$, corresponding to an absolute colour difference of only $0.06$ mags at $3300K$, while at near infrared wavelengths the colour differences are negligible.  We do find that the Johnson $B-V$ colour does show some sensitivity to reddening, again at the coolest extreme of the temperature range, but only $\sim~0.04$ mags.  We have applied the same fits with the solar metallicity $15M_{\odot}$ to the observed datasets and find that the differences in quality of the fits is small when compared to, for example, the effect of metallicity; so we do not consider the $15M_{\odot}$ models further. 
%%%%%%%%%%%%%%%%%%%%%%%%%%%%%%%%%%%%%%%%%%%%%%%%%%%%%%%%%%%%%%%%%%%%
%TABLE MODEL SELECTION - tab:ana:sedfit
%%%%%%%%%%%%%%%%%%%%%%%%%%%%%%%%%%%%%%%%%%%%%%%%%%%%%%%%%%%%%%%%%%%%
\begin{table*}
\caption{\label{tab:ana:sedfit} The results of SED fits to MARCS synthetic spectra to the observations of the progenitor of SN~2008bk.}
\begin{tabular}{clccccc}
\hline\\
[-1.5ex]
  $\log (Z/Z_{\odot})$     &      & $R_{V}=2$ & $R_{V}=3.1$ & $R_{V}=4$ & $R_{V}=5$ & $R_{V}=3.1+\mathrm{Sil}$ \\[1ex] \hline \\ [-1.5ex]
   
  & & \multicolumn{5}{c}{$g'r'i'BVIYJHK_{S}$ $(N=8, M=2)$}\\  
        \cline{3-7}\\[1.0ex] 
     -0.25 &  $K$       &$-29.4\pm0.4$& $0.0\pm0.0$& $14.2\pm0.4$& $21.0\pm0.4$& $-33.3\pm0.6$\\ 
     & $\chi^{2}$        &$36.0$&  $22.0$& $15.2$&  $10.6$&  $31.47$\\ [2.5ex]
    0.0  &  $K$       &$-33.5\pm0.4$& $-4.2\pm0.4$& $9.7\pm0.4$& $16.8\pm0.4$& $-39.8\pm0.6$\\ 
     & $\chi^{2}$        &$38.1$&  $23.9$&  $17$&  $12.3$& $34.0$\\
   [1ex] \hline \\ [-1.5ex]  
 
   & & \multicolumn{5}{c}{$g'r'i'BVIJHK_{S}$ $(N=7, M=2)$}\\  
         \cline{3-7}\\[1.0ex] 
  -0.25  & $K$        &  $-29.2\pm0.4$& $0.0\pm0.0$& $13.4\pm0.4$& $19.8\pm0.4$& $-33.5\pm0.4$\\ 
        & $\chi^{2}$  & $35.5$&  $21.7$&  $15$&  $10.2$&  $30.7$\\ [2.5ex]
    0.0 & $K$        &$-34.0\pm0.4$& $-4.6\pm0.4$& $8.9\pm0.4$& $16.6\pm0.4$& $-37.7\pm0.4$\\ 
        & $\chi^{2}$ &$37.7$&  $23.7$&  $16.8$&  $12$&  $33.4$\\
   [1ex] \hline \\ [-1.5ex]
      & & \multicolumn{5}{c}{$r'i'BVIJHK_{S}$ $(N=6,M=2)$}\\ 
      \cline{3-7}\\[1.0ex] 
    -0.25 & $K$        &$-5.9\pm0.4$& $0.0\pm0.0$& $1.4\pm0.4$& $-0.7\pm0.4$& $-13.1\pm0.5$\\ 
     & $\chi^{2}$        &$8.7$&  $5.4$&  $5.3$& $5.6$& $8.3$\\[2.5ex]
     0.0 &  $K$       &$-9.4\pm0.4$& $-1.8\pm0.4$& $-1.7\pm0.4$& $-3.3\pm0.4$& $-16.5\pm0.5$\\ 
     & $\chi^{2}$        &$10.1$&  $6.9$& $6.8$& $6.7$& $10.0$\\
   [1ex] \hline \\ [-1.5ex]
         & & \multicolumn{5}{c}{$r'i'BVIYJHK_{S}$ $(N=7, M=2)$}\\  
               \cline{3-7}\\[1.0ex] 
    -0.25 &  $K$       &$-6.6\pm0.4$& $0.0\pm0.0$& $-1.0\pm0.4$& $-2.1\pm0.4$& $-14.6\pm0.5$\\ 
     & $\chi^{2}$        &$9.02$& $5.8$&  $5.8$& $6.1$&  $8.8$\\ [2.5ex]
     0.0 &  $K$    &$-9.9\pm0.4$& $-2.6\pm0.4$& $-3.2\pm0.4$& $-11.6\pm0.4$& $-17.1\pm0.5$\\ 
     & $\chi^{2}$        &$10.3$&  $7.1$&  $7.1$&  $7.7$&  $10.3$ \\
 [1ex] \hline \\ [-1.5ex]  
\end{tabular}
\end{table*} 
%%%%%%%%%%%%%%%%%%%%%%%%%%%%%%%%%%%%%%%%%%%%%%%%%%%%%%%%%%%%%%%%%%%%

To explore the parameters of the progenitor we consider the dataset composed of all the photometric detections and non-detections except the Gemini GMOS $g^{\prime}$ observation.  Posterior samples, generated during the determination of the Bayesian evidence for the SED fitting process, were used to estimate the median and $68\%$ probability interval for each of the model fits to the observed SED of the progenitor.  The best fit parameters, for different assumptions of metallicity and reddening law, are presented in Table \ref{tab:ana:par}. As expected, increases in the value of $R_{V}$ yield lower reddenings and higher temperatures for the progenitor.  For each reddening law, the assumption of solar metallicity results in a slightly hotter, less reddened progenitor.  The inclusion of a Silicate dust component, under the assumption of low interstellar reddening (see above), yields not only a poor solution (see Table \ref{tab:ana:sedfit}) but also low values of $\tau$.

\begin{figure*}
\includegraphics[width=15cm,angle=270]{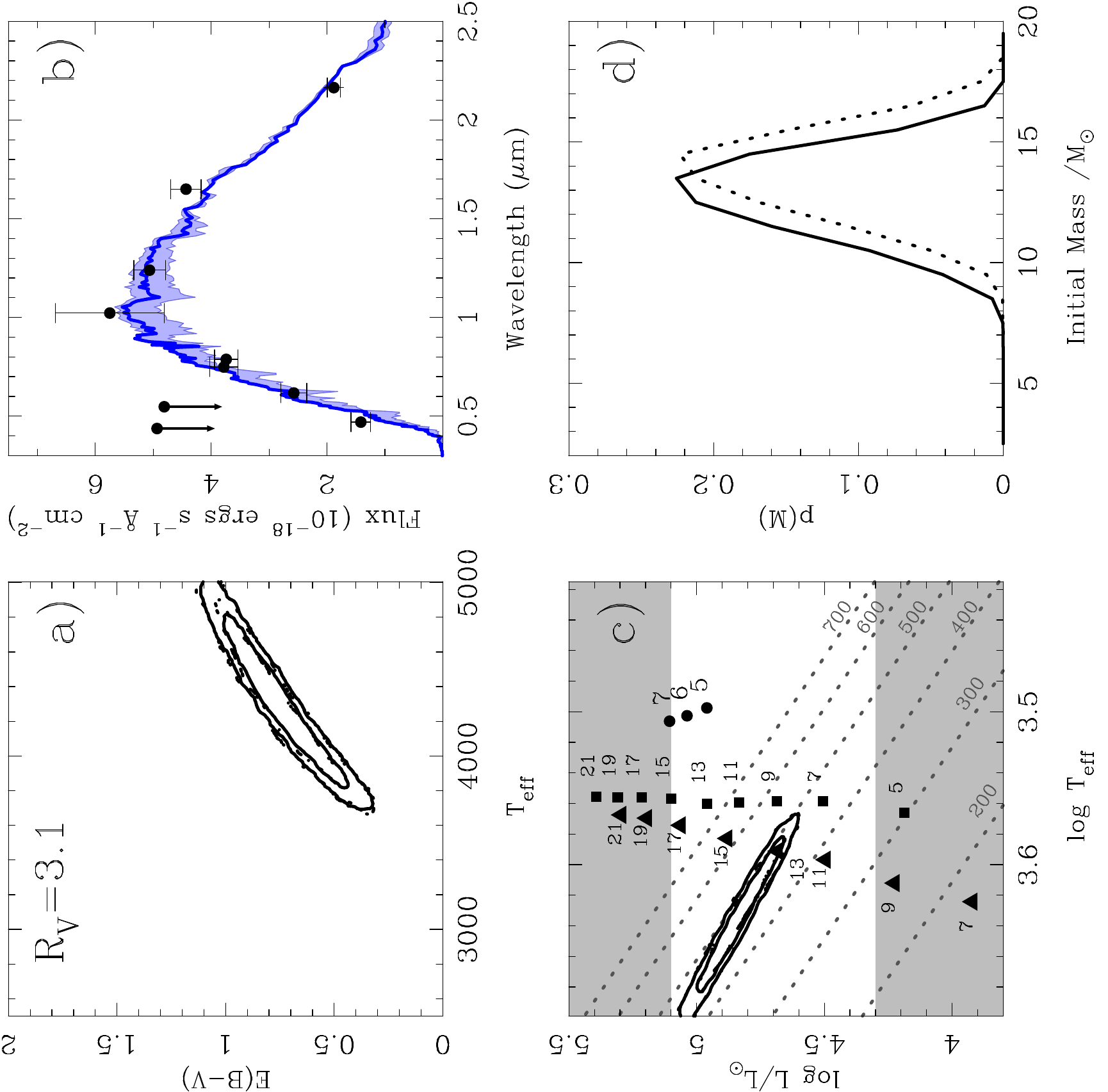}
\caption{The derived properties of the progenitor of SN~2008bk,
assuming $R_{V}=3.1$ for $\log (Z/Z_{\odot}) = -0.25$ and $0.0$.  {\it a)}  The joint posterior distribution for reddening $E(B-V)$ and temperature $T_{eff}$.  The solid and dashed contours correspond to metallicities of $\log (Z/Z_{\odot}) = -0.25$ and $0.0$, respectively.  The contours contain $68\%$ and $95\%$ of the probability for each model fit.  {\it b)}  The observed photometry, and detection limits, of the progenitor compared to the best fit SED model (heavy blue line).  The light blue shaded region are the SED boundaries corresponding to the $68\%$ probability interval.  {\it c)}  The position of the progenitor on the HR diagram.  The meaning of the contours is the same as above.  The shaded regions indicate locations on the HR diagram that are inconsistent with the measured $K_{S}$-band brightness.  Over-plotted are the points from LMC metallicity STARS stellar evolution models, with initial masses as labeled, corresponding to the positions of the end of He-burning ($\blacktriangle$), the termination point of the models ($\blacksquare$) and the termination point for those stars that become SAGB stars ($\bullet$).  The dashed grey lines show lines of constant radius on the HR diagram. {\it d)}  Mass probability density distributions for the progenitor of SN~2008bk, for $\log (Z/Z_{\odot}) = -0.25$ (solid) and $0.0$ (dashed).}
\label{fig:ana:hrd31}
\end{figure*}

For the LMC and solar metallicities, assuming $R_{V}=3.1$, the best fit solution to the observed data (excluding the $g^{\prime}$ observation) is presented on Figure \ref{fig:ana:hrd31}.  From this analysis, the joint posterior distribution for the temperature and reddening of the progenitor (Figure \ref{fig:ana:hrd31}a) shows the standard degeneracy, with SEDs with increasing temperatures requiring corresponding increases in reddening to match the observed colours of the progenitor object.  The contours for the solar and LMC metallicities almost completely overlap, although there are some minor differences, and this is reflected in slight differences seen in Table \ref{tab:ana:sedfit} for model fits with different metallicities.  The model SED (Figure \ref{fig:ana:hrd31}b) provides a good match to the observations, although we note the $H$-band magnitude appears to lie above the $68\%$ interval by $1\sigma$.  The model fits are dominated by the IR photometry and, hence, the best SED fits consistently underestimate the observed $g^{\prime}$ flux.
 
The luminosity and temperature constraints for the progenitor on the HR diagram (Figure \ref{fig:ana:hrd31}c) show the degeneracy between the temperature and the derived bolometric luminosity (through the bolometric and extinction/reddening corrections).  The luminosity-temperature contours only slightly overlap the model points for the end of He-burning for a $13M_{\odot}$ star, but do not extend to cooler temperatures consistent with the actual end points of the STARS stellar evolution models \citep{eld04}, at which core-collapse is expected to take place.  The slope of the contours lies approximately parallel to the direction of constant radius of the HRD, with the progenitor's radius constrained to $\sim400-500R_{\odot}$ for both the LMC and solar metallicity models.  The luminosity  derived from the SED fitting analysis is consistent with the luminosity constraints derived from the $K_{S}$-band brightness.

We adopt the same technique presented by \citet{2013arXiv1302.7152M} to determine the mass probability density function for the LMC and solar metallicity SED fits with $R_{V}=3.1$.  \citeauthor{2013arXiv1302.7152M} assume that, for a given luminosity, the progenitor has a uniform probability of having a mass between the maximum mass for a star to end He-burning at that luminosity and the minimum mass for a star to terminate (i.e. the final end point of the model) at that luminosity.  Using the luminosity probability distributions derived above, in comparison with the He-burning and model end points presented by \citet{2008arXiv0809.0403S}, we derive the probability density function for the progenitor having a specific initial mass (see Figure \ref{fig:ana:hrd31}d).  The mass estimates for the two metallicities considered are similar, although the solar metallicity yields a slightly higher mass.  For $R_{V}=3.1$, we derive $M_{init}=12.9^{+1.6}_{-1.8}M_{\odot}$ and 
$13.8^{+1.6}_{-1.8}M_{\odot}$ for LMC and solar metallicities respectively.

%%%%%%%%%%%%%%%%%%%%%%%%%%%%%%%%%%%%%%%%%%%%%%%%%%%%%%%%%%%%%%%%%%%%
%FIGURE HRD PANEL - fig:ana:hrd31
%%%%%%%%%%%%%%%%%%%%%%%%%%%%%%%%%%%%%%%%%%%%%%%%%%%%%%%%%%%%%%%%%%%%
%%%%%%%%%%%%%%%%%%%%%%%%%%%%%%%%%%%%%%%%%%%%%%%%%%%%%%%%%%%%%%%%%%%%
%TABLE BEST FIT PARAMETERS - tab:ana:par
%%%%%%%%%%%%%%%%%%%%%%%%%%%%%%%%%%%%%%%%%%%%%%%%%%%%%%%%%%%%%%%%%%%%
\begin{table*}
\caption{\label{tab:ana:par}  The best fit parameters for the progenitor of SN~2008bk.}
\begin{tabular}{lcccccc}
\hline\hline
$R_{V}$               & $\log (Z/Z_{\odot})$ & $E(B-V)$                  & $T_{eff}\, (K)$      & $\log (L/L_{\odot})$  & $R/R_{\odot}$   & $M/M_{\odot}$ \\[1ex]
\hline 
 2                     &-0.25                & $0.98^{+0.22}_{-0.23}$    & $4025^{+215}_{-180}$ & $4.71\pm0.07$         &$465\pm14$       &$11.3\pm1.5$\\[1ex]
                       &0.0                  & $1.00\pm0.24$             & $4030^{+235}_{-190}$ & $4.72\pm0.07$         &$470^{+13}_{-16}$&$12.4\pm1.7$\\[1.5ex]
\marktopleft{c1} 3.1   &-0.25                & $0.77^{+0.17}_{-0.21}$    & $4330^{+330}_{-335}$ & $4.84\pm0.11$         &$470\pm16$       &$12.9^{+1.6}_{-1.8}$  \\[1ex]
                       &0.0                  & $0.78^{+0.17}_{-0.21}$    & $4325^{+340}_{-330}$ & $4.85\pm0.11$         &$475\pm16$       &$13.8^{+1.6}_{-1.8}$ \markbottomright{c1}\\[1.5ex]
 4                     &-0.25                & $0.63^{+0.12}_{-0.17}$    & $4450^{+350}_{-390}$ & $4.88\pm0.11$         &$475\pm16$       &$12.9^{+1.6}_{-1.8}$ \\[1ex]
                       &0.0                  & $0.64^{+0.13}_{-0.18}$    & $4475^{+360}_{-420}$ & $4.90^{+0.11}_{-0.14}$&$475\pm17$       &$15.5^{+1.7}_{-2.0}$ \\[1.5ex]
 5                     &-0.25                & $0.50^{+0.10}_{-0.15}$    & $4480^{+365}_{-415}$ & $4.91^{+0.12}_{-0.15}$&$475\pm14$       &$13.9^{+1.6}_{-2.2}$\\[1ex]
                       &0.0                  & $0.57^{+0.05}_{-0.08}$    & $4775^{+170}_{-295}$ & $5.00^{+0.05}_{-0.09}$&$465\pm12$       &$15.5\pm1.4$\\[1.5ex]
$3.1+\mathrm{Sil}$     &-0.25                & $\tau=3.3^{+0.8}_{-0.9}\,{\dagger}$& $3890^{+150}_{-130}$ & $4.65\pm0.04$         &$465\pm15$       &$10.5\pm1.4$\\[1ex]
                       &0.0                  & $\tau=3.3^{+0.8}_{-1.0}\,^{\dagger}$& $3880^{+160}_{-130}$ & $4.65\pm0.05$         &$465\pm16$       &$11.7\pm1.6$\\[1ex]
\hline\hline
\end{tabular}
\end{table*}
%%%%%%%%%%%%%%%%%%%%%%%%%%%%%%%%%%%%%%%%%%%%%%%%%%%%%%%%%%%%%%%%%%%%
\section{Discussion \& Conclusions}
\label{sec:disc}

\subsection{The nature of the progenitor of SN~2008bk}
\label{sec:disc:prog}
We have measured the properties of the progenitor of Type IIP SN 2008bk in pre-explosion observations coupled with late-time observations, using the same telescopes and instruments, in which the progenitor  (and the SN) are no longer observed.  The absence of the pre-explosion star in the late-time images confirms the original identification of this star as the progenitor \citep{2010arXiv1011.5494M}.  Using image subtraction techniques we have determined the infrared brightness of the progenitor to much higher precision than allowed by using the pre-explosion observations alone.  With the benefit of late-time {\it HST} observations we have estimated the residual SN flux contamination to our template subtraction analysis, as well as demonstrated that there is little contamination at IR wavelengths from surrounding stars.   In the optical, the complex and crowded nature of the region hosting the SN most probably leads to overestimates of the progenitor brightness from photometry of the pre-explosion observations alone.  With the revised SED, constructed from improved photometry, we have shown the progenitor has a relatively high mass for a Type IIP SN progenitor: $\sim 13M_{\odot}$. The progenitor is found to have suffered large reddening and, correspondingly, had a high temperature.  

Our method for determining the masses of the progenitors is different to those previously used, in particular the technique adopted by \citet{2008arXiv0809.0403S}.  We believe our methodology, presented by \citet{2013arXiv1302.7152M}, is more robust because it correctly couples the uncertainties on the observed luminosities with the uncertainty on what luminosity, between the end of core He-burning and the model end point, an RSG progenitor may actually be observed at.  The consideration of the various parameters underlying the SED fits (such as a reddening law and metallicity), through the use of a complete family of SED models, means that we can comprehensively assess systematic effects.  The improved precision of our progenitor photometry, in conjunction with the systematic comparison of the observations with model RSG SEDs, leads to a more precise and confident mass estimate for the progenitor of SN~2008bk than has been previously achieved for any other Type IIP SN from pre-explosion images.  In section \ref{sec:disc:prev}, we compare the results of this new analysis with the previous estimates of the progenitors properties for SN~2008bk.

As found previously by \citeauthor{2013arXiv1302.7152M} for other RSG progenitor detections, this new approach yields four interesting results concerning the progenitor of SN~2008bk: 1) the initial mass probability density function extends up to slightly higher masses than suggested by the previous analysis; 2) there is evidence for the presence of significant dust affecting the progenitor, that is not perhaps observed in post-explosion observations of the SN; 3) the effect of this dust is well described by a standard reddening law consistent with ISM dust;  and 4) the derived temperature of the progenitor is higher than would be expected if the progenitor was a late M-supergiant.

\subsection{Results from previous analyses}
\label{sec:disc:prev}
Previous analyses of the progenitor of SN~2008bk have been presented by \citet{2008ApJ...688L..91M},\citet{2012AJ....143...19V} and \citet{2013arXiv1302.2674D}.   The analysis of \citeauthor{2013arXiv1302.2674D} is based on the photometric measurements presented by \citeauthor{2008ApJ...688L..91M}, while \citeauthor{2012AJ....143...19V} present a reanalysis of the same IR data used by \citeauthor{2008ApJ...688L..91M} supplemented with optical detections of the progenitor.  In this paper, we have reanalysed all the previously presented pre-explosion datasets, with more accurate calibrations based on late-time observations of the same fields acquired under photometric conditions.

A comparison of the photometry derived here and those of the previous studies is presented in Table \ref{tab:res:phot}.  The VLT IR photometry common to both previous studies shows some agreement and disagreement with our photometry.  The $J$ and $K_{S}$ magnitudes we measure from the ISAAC data are almost identical to those presented by \citeauthor{2008ApJ...688L..91M}, whilst being $0.2-0.3$ mags fainter than measured by \citeauthor{2012AJ....143...19V}.  Conversely, our $H$-band measurement from the HAWKI data agrees with \citeauthor{2012AJ....143...19V}, but is $0.26$ mags fainter than measured by \citeauthor{2008ApJ...688L..91M}. 

The key measurements affecting the analysis of the SED are the Cousins $I$ and Sloan $i^{\prime}$ brightnesses.  We find our measurements of the $I$-band brightness to be almost identical to those of \citeauthor{2008ApJ...688L..91M}, although slightly fainter due to the  removal of contaminating flux from surrounding stars through the use of template subtraction with late-time images.  The measurement by \citeauthor{2012AJ....143...19V} is significantly higher, however the full origin of this discrepancy is difficult to identify as they bootstrapped their $i^{\prime}$ photometry to Cousins $I$ (see below); and the same is the case for comparing our $g^{\prime}$ and $r^{\prime}$ photometry with their $VR$ photometry.     Given the apparent consistency we achieve independently between our VLT FORS Cousins $I$ and Gemini $i^{\prime}$ photometry, we believe that the lower value of the $I$ band brightness is correct.  Even though we have established the stability of observing under photometric conditions with Gemini GMOS, over the period in which the late-time observations were conducted, the lack of photometric standards observed on the same night as late-time $g^{\prime}$ and $r^{\prime}$ observations raises some concern about the size of the uncertainties estimated for these two measurements.\\

Using late-time $HST$ observations we have established that the degree of contamination of the pre-explosion photometry from {\it unresolved} blended stars was smallest at optical wavelengths.  At IR wavelengths the pre-explosion photometry of the source at the SN position is dominated by the bright progenitor, compared to other, fainter sources nearby (see Table \ref{tab:res:hstphot}).  It is not surprising that, at IR wavelengths, the photometry of the progenitor in the pre-explosion images with and without late-time image template subtractions should be similar.  As evident from the optical ground-based images, the region in which SN 2008bk is located is crowded, such that PSF photometry of a single object may be suboptimal in the presence of bright, yet resolved neighbours.  We specifically note the discrepancy between our photometry of the pre-explosion $i^{\prime}$ source, from the pre-explosion image alone, and the final magnitude we derive for the progenitor with the benefit of template subtraction.  As our $g^{\prime}$ and $r^{\prime}$ photometry was not conducted with the benefit of template subtraction, due to the SN still being bright at these wavelengths, these measurements maybe be overestimates due to the effect of crowding (as shown by the $i^{\prime}$ photometry).  This may partially explain the apparent discrepancy between the model SEDs and the $g^{\prime}$ photometry, and may cast doubt over the veracity of the $r^{\prime}$ measurement.  The differences between our analysis and the previously published studies highlights the importance of acquiring late-time observations of SNe with identified progenitors: to not only confirm the disappearance of the progenitor, but to also achieve both precise and accurate photometry of the progenitor object.

\begin{figure*}
\includegraphics[width=8cm,angle=270]{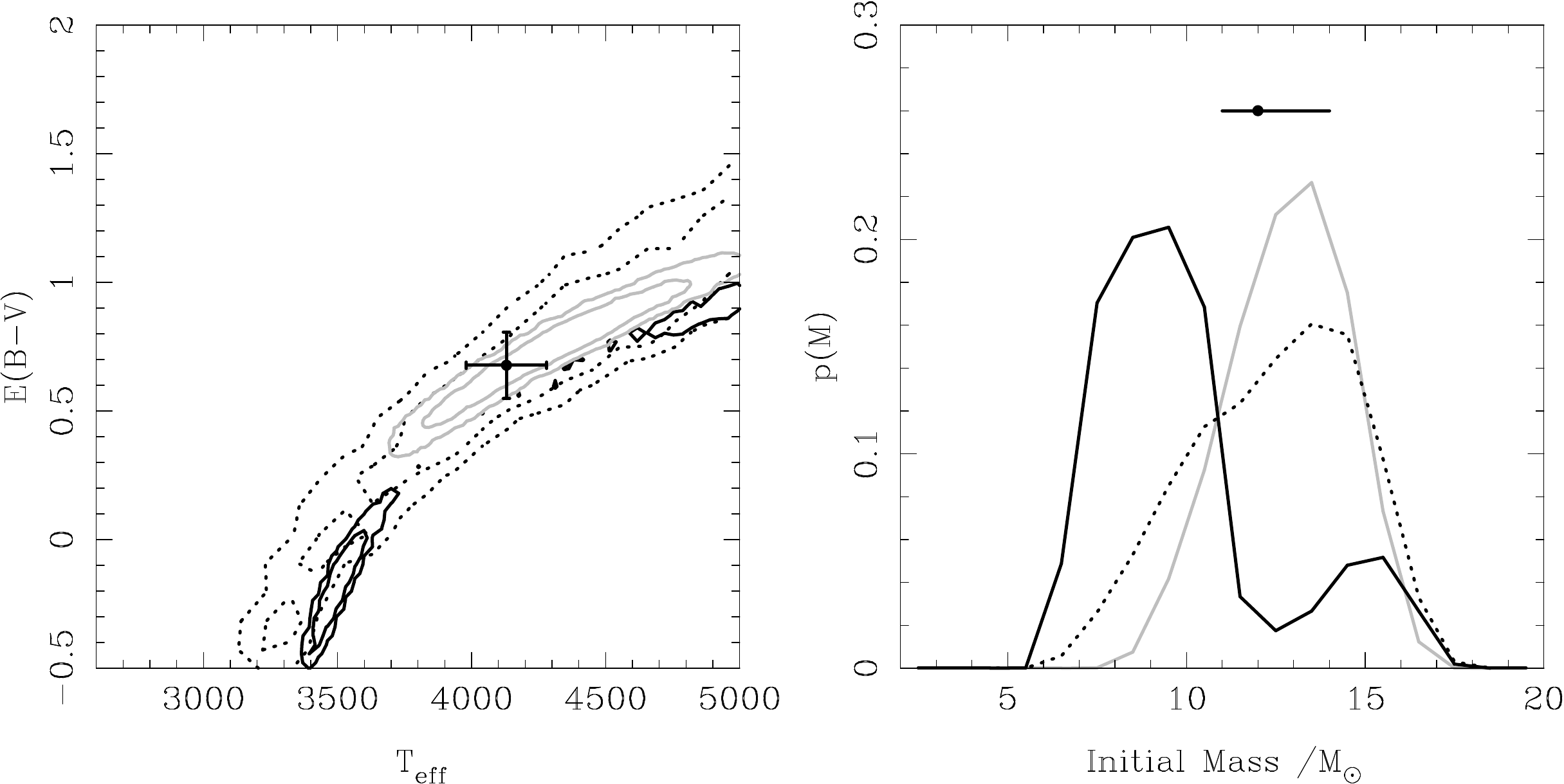}
\caption{The properties of the progenitor derived from analysis of the photometry presented by \citet{2008ApJ...688L..91M} and \citet{2012AJ....143...19V}. {\it Left)}  The joint posterior distributions for $E(B-V)$ and $T_{eff}$ given the datasets of \citeauthor{2012AJ....143...19V} (solid contours) and \citeauthor{2008ApJ...688L..91M} (dotted contours).  The derived reddening and temperature from the analysis of \citet{2013arXiv1302.2674D} is indicated by the point. {\it Right)} Mass probability density functions from analyses of the previous datasets (following the scheme of the left panel). For comparison our result, presented in Figure \ref{fig:ana:hrd31}, is shown in grey.}
\label{fig:disc:other}
\end{figure*}

Using the previously reported photometry of \citeauthor{2008ApJ...688L..91M} and \citeauthor{2012AJ....143...19V}, we conducted the same analysis as presented in Section \ref{sec:ana:sed} on their datasets.  The derived properties of the progenitor given these previous measurements are presented on Figure  \ref{fig:disc:other}.  

The reanalysis of the \citeauthor{2008ApJ...688L..91M} data by \citet{2013arXiv1302.2674D} is consistent with the result derived here with an expanded dataset.  \citeauthor{2013arXiv1302.2674D} derive a mass of $12^{+2}_{-1}M_{\odot}$,  higher than the mass derived by \citeauthor{2008ApJ...688L..91M} using the same dataset.  The two studies, however, used fundamentally different underlying SEDs to estimate the progenitor parameters.  Like us, \citeauthor{2013arXiv1302.2674D} compared this observed photometry with MARCS SEDs, whereas \citeauthor{2008ApJ...688L..91M} compared their photometry with the empirical RSG colour scheme of \citet{1985ApJS...57...91E}.  Our reanalysis of the \citeauthor{2008ApJ...688L..91M} photometry, as published, and the similarity with the result obtained by \citeauthor{2013arXiv1302.2674D}, might suggest the fundamental difference in the result is due to the differences in the choice of underlying SEDs; however, as discussed below, we do not believe this is the case.  We note that the parameters of the progenitor, as constrained by the \citeauthor{2008ApJ...688L..91M} dataset, do not completely resolve the degeneracy between temperature and reddening, as indicated by the barely closed contours on Figure \ref{fig:disc:other}, although they favour the high temperature solution found in our analysis. 

In their original analysis, \citeauthor{2008ApJ...688L..91M} favoured a low mass progenitor ($M_{init}=8.5 \pm 1M_{\odot}$), assuming an M4 spectral type.  They also noted, however, that the observed SED might also be accommodated with a progenitor with earlier spectral type and higher reddening (going so far as to consider a G-type yellow supergiant).  For an M0 supergiant progenitor, with $T_{eff}=3750K$ and $A_{V}=3$, they suggested a mass as high as $11\pm2M_{\odot}$, which is approaching the value determined here (although with higher reddening).

 As for the progenitor of SN~2012aw \citep{2012arXiv1204.1523F}, the \citeauthor{2008ApJ...688L..91M} analysis was based on four photometric measurements.  It is clear from our analysis of the progenitor of SN~2008bk, with the expanded dataset, that $>4$ photometric points are required to adequately constrain the progenitor SED and break the degeneracies between temperature and reddening.
%SVD

The analysis by \citeauthor{2012AJ....143...19V} suggested a low reddening towards the progenitor, however our reanalysis, despite using the same synthetic spectra, suggests a different picture.  It is evident from Figure \ref{fig:disc:other}, for their photometry, that the bulk of the probability in the reddening/temperature plane lies at negative reddening (driven by the $g^{\prime}$ measurement).  The apparent need for negative reddening to provide an adequate fit to the theoretical SEDs may suggest: 1) the models are incompatible with the observed data or 2) the data are incompatible with the theoretical models (based on our photometry, we favour the latter).  The low reddening quoted by \citeauthor{2012AJ....143...19V} is in actuality a "grid edge" effect, based on the assumption that $E(B-V) \geq 0$.  Such an assumption could be introduced into our analysis of their dataset, in the form of a prior, however we note that: 1) the corresponding Bayesian evidence would be even lower than the value determined for the complete parameter space;  and 2) the most probable solution would then be with high reddening and temperature.  When including the negative reddening solution, the \citeauthor{2012AJ....143...19V} dataset yield a bimodal solution; with the negative reddening solution supporting a low mass progenitor (consistent with their estimate of $8-8.5M_{\odot}$) and the other probability peak consistent with a high mass progenitor (at the extreme of the mass range determined by our analysis).

\subsection{Implications for RSGs}
\label{sec:disc:rsgs}
The temperature inferred for the progenitor of SN~2008bk is higher than generally reported for RSGs.  We have principally compared our observations with the Cambridge STARS code \citep{eld04}, which generally predicts RSGs to have cooler temperatures.  The inclusion of rotation into the models of \citet{2012A&A...537A.146E}, for $\log (Z/Z_{\odot})\sim -0.15$, does not cause a significant change in the predicted temperatures of RSGs, when compared with their non-rotating models (at the luminosity at which we have observed the progenitor of SN~2008bk).   \citet{2008arXiv0809.0403S} presented a compilation of the temperatures at the endpoints, for a range of initial masses, of a variety of different stellar evolution models (see their Fig. 3), which are all significantly cooler than the temperature we measure here.

Our inferred temperature agrees with the temperature-colour relation presented by \citet{2006ApJ...645.1102L}, where for the progenitor of SN~2008bk we find $(V-K)_{0}=3.04$.  Following the spectral type sequence of \citeauthor{2006ApJ...645.1102L}, the progenitor of SN~2008bk would be considered to have spectral type K1; significantly earlier than the canonical prediction of M-supergiants as the progenitors of Type IIP SNe, and just as hot as the progenitor of the Type IIb SN~1993J \citep{alder93j,maund93j}. 

An important concern, therefore, is that either the MARCS SEDs yield solutions that are too hot or stellar evolution models underestimate the temperature of the progenitors.  We note, however, that the intrinsic colours for the progenitor (for the best fit solution) are also consistent with the colour sequence of \citet{1985ApJS...57...91E}; and, following their spectral type sequence, we again find the progenitor is consistent with an early K-supergiant.  While our result is dependent on MARCS SEDS, we note that \citet{2012MNRAS.419.2054W} derived a similar mass but using the BaSeL standard stellar library \citep[][and references therein]{2002A&A...381..524W}.  Conversely, the temperatures predicted by stellar evolution models may be "tuned" through the assumptions of different opacities and mixing lengths.

\citet{2013arXiv1302.2674D} recently compared temperature estimates for RSGs derived using SED fitting techniques (similar to those used here) and the strengths of the TiO bands.  \citeauthor{2013arXiv1302.2674D} found that temperatures determined from the strength of the TiO bands  are generally lower ($\sim 400-500K$) than those derived from IR photometry; and, furthermore, the strength of the TiO bands were observed to be correlated with the luminosity of the RSG.  In light of the new analysis by \citeauthor{2013arXiv1302.2674D}, we find that the temperature for the progenitor of SN~2008bk is not unusual.  In previous analyses, with single detections or upper limits, we have had to assume the temperature/spectral type of a cool RSG\citep[see e.g][]{2008arXiv0809.0403S}; such that if the temperatures were underestimated, the corresponding bolometric luminosities were overestimated (making the RSG problem even worse).  Our recent reanalysis of the pre-explosion observations of the progenitors of SNe 2003gd and 2005cs have, similarly, resulted in hotter temperature constraints (in the similar temperature range as the progenitor of SN 2008bk; \citealt{2013arXiv1302.7152M}).

\subsection{Implications for Type IIP SNe}
\label{sec:disc:iip}
Based on the cool temperatures originally assumed for the progenitors of SNe 2005cs \citep{2005astro.ph..7502M,2005astro.ph..7394L} and 2008bk \citep{2008ApJ...688L..91M} (and the properties of the subsequent SNe), \citet{2009ApJ...705L.138P} suggested the progenitors may have been Super Asymptotic Giant Branch (SAGB) stars exploding as "electron capture" SNe \citep{nom84}, rather than due to the collapse of an iron core.  Our revised temperature for the progenitor of SN~2008bk clearly rules it out as being an SAGB star; and the new initial mass estimate suggests the progenitor was capable of producing an Fe-core at the end of its life.    

The progenitor mass provides a key test of hydrodynamical models of SN evolution (specifically the behaviour of the light curve).  For SN~2005cs there was a clear discrepancy between the progenitor mass determined from pre-explosion observations \citep[$7-12M_{\odot}$][]{2005astro.ph..7502M,2005astro.ph..7394L} and the hydrodynamical mass \citep[$\sim 18M_{\odot}$][]{2008A&A...491..507U}.  The recent reanalysis of the pre-explosion observations of SN~2005cs, using late-time observations \citep{2013arXiv1302.7152M} suggested the mass of the progenitor of SN~2005cs might be increased slightly ($\sim 1M_{\odot}$), but not enough to resolve the discrepancy.  The increase in mass from our new analysis of the progenitor of SN~2008bk might go some way to resolve the discrepancy between the our progenitor masses and the hydrodynamical masses.  Similarly, the revised mass estimate for this progenitor may suggest previous mass estimates for the progenitors of other Type IIP SNe have been underestimated.  This may have contributed to the RSG problem \citep{2008arXiv0809.0403S}; however if the small increase in mass we have found for the progenitor of SN~2008bk were equally applicable to all progenitors, it would not completely eliminate the RSG problem.  Although SN~2008bk was identified as a low-luminosity (low velocity, low Ni mass) Type IIP SN, the mass we derive for the progenitor is higher than previously determined for the progenitor of the similar SN~2005cs\citep{2013arXiv1302.7152M}.

\citet{2012MNRAS.420.3451M} analysed the nebular spectrum of SN~2008bk, and showed that the observed line fluxes were consistent with a model spectrum arising from  a $12M_{\odot}$ progenitor, matching the result found here using pre-explosion observations.  Recently, \citet{2012A&A...546A..28J}
showed these line fluxes, in particular $[ O\, I]$, could be used to establish the progenitor mass to within $\sim 3M_{\odot}$. 

We note that the radius constraint for the progenitor of SN~2008bk presented here is at the lower end of the radius range of observed RSGs \citep[$500-1500R_{\odot}$; see e.g.][]{masslmc03,2007ApJ...667..202L,2006ApJ...645.1102L}.  \citet{2013MNRAS.433.1745D} found, however, that RSGs with larger radii ($600-1100R_{\odot}$) would produce SN light curves with a $U$-band plateau before fading.  To match the observed colour evolution of Type IIP SNe, they suggested the progenitors must be more compact stars; with radii similar to the radius we find for the progenitor of SN~2008bk. 

\subsection{The role of dust for the progenitors of Type IIP SNe}
\label{sec:disc:dust}
The dust responsible for the reddening of the progenitor of SN~2008bk most likely has a similar composition to interstellar dust (with an average $R_{V}=3.1$ reddening law).  The large reddenings inferred to nearby {\sc Hii} regions and the observation of a significant light echo in HST observations \citep{2013arXiv1305.6639V}, may suggest that the bulk of the reddening affecting the progenitor of SN~2008bk was in the ISM and not local to the star itself.  The large amount of dust affecting the SED of the progenitor of SN~2008bk is, however, in conflict with the low reddening derived from the observations of SN 2008bk itself, in the analysis of \citet{2012AJ....143...19V}; and is higher than estimated by \citet{2008ApJ...688L..91M} for their low mass solution. 

 Although dust has been suggested as a possible explanation for the RSG problem \citep{2012MNRAS.419.2054W,2012ApJ...759...20K}, our analysis of the progenitor of SN~2008bk (and previous findings \citealt{2013arXiv1302.7152M}) have not indicated that any CSM dust affecting the progenitors of Type IIP SNe have a different composition or reddening law compared to the ISM dust.  While \citep{2012ApJ...759...20K} suggested Silicate or Graphite dust might explain the observed SED of the progenitor of SN~2012aw, the available data for that progenitor is insufficient to rule out ordinary ISM dust.  The conflicting reported reddening estimates to SN~2008bk itself (and towards objects in the vicinity in NGC 7793), derived from post-explosion observations, do not clarify if a large volume of dust in the CSM was destroyed by the SN \citep{1983ApJ...274..175D,1986A&A...155..291P}.

In the process of analysing the HST data, we note that we find disagreement between our photometry (presented in Table \ref{tab:res:hstphot}) and the photometry presented by \citet{2013arXiv1305.6639V}.  In particular we find their $V-I$ colours to be too blue, being inconsistent with the optical and infrared colours predicted by MARCS spectra.    
%implications for EC and normal Type IIP SNe.
%%%%%%%%%%%%%%%%%%%%%%%%%%%%%%%%%%%%%%%%%%%%%%%%%%%%%%%%%%%%%%%%%%%%
%FIGURE PREVIOUS RESULTS - fig:disc:other
%%%%%%%%%%%%%%%%%%%%%%%%%%%%%%%%%%%%%%%%%%%%%%%%%%%%%%%%%%%%%%%%%%%%
%%%%%%%%%%%%%%%%%%%%%%%%%%%%%%%%%%%%%%%%%%%%%%%%%%%%%%%%%%%%%%%%%%%%
\section*{Acknowledgments}
The research of JRM is funded through a Royal Society University Research Fellowship.

\bibliographystyle{apj}
%\bibliography{/Users/justyn/Documents/Bibliography/main}

\end{document}